\documentclass[11pt]{article}
\usepackage{mathrsfs}
\usepackage{amsfonts}
\usepackage{amsfonts}
\usepackage{mathrsfs}
\usepackage{amssymb}
\usepackage{hyperref,color}
\usepackage{exscale,relsize}

\usepackage{amsmath,amsthm}
\numberwithin{equation}{section}

\newcommand{\eqn}[1]{\begin{equation*}#1\end{equation*}}
\newcommand{\eql}[2]{\begin{equation}\label{#1}#2\end{equation}}
\newcommand{\eqnarrayl}[2]{\begin{eqnarray}\label{#1}#2\end{eqnarray}}
\newcommand{\eqnarrayn}[1]{\begin{eqnarray*}#1\end{eqnarray*}}

\def\al{\alpha}
\def\be{\beta}
\def\h{\hbar}
\def\cur{\mathcal{K}_p}

\allowdisplaybreaks[1]

\author{\bf Peng Zhao$^1$,~Engui Fan$^1
\footnote{Corresponding
author and  e-mail address:
      faneg@fudan.edu.cn}$,~ Yu Hou$^1$}
\date{ \small{$^1$ School of Mathematical Science, Fudan University, Shanghai
200433, P.R. China}}
\title{\bf The Algebro-Geometric Solutions for the  Ruijsenaars-Toda Hierarchy}
\begin{document}
\maketitle

\begin{abstract}
We provide a detailed treatment of  Ruijsenaars-Toda (RT) hierarchy with special
emphasis on its the theta function representation of all algebro-geometric solutions.
The basic tools involve
hyperelliptic curve  $\mathcal{K}_p$ associated with the Burchnall-Chaundy polynomial,
Dubrovin-type
equations for auxiliary divisors and associated trace formulas. With the help of
a foundamental meromorphic
function $\phi$, Baker-Akhiezer vector $\Psi$ on $\mathcal{K}_p$,
the complex-valued algebro-geometric solutions
of RT hierarchy are derived.
\end{abstract}

\section{Introduction}

~~~~Nonlinear integrable lattice systems have been studied extensively in relation with
various aspects and they usually possess rich mathematical structure such as Lax pairs,
Hamilton structure, conservation law, etc.
The Toda lattice is one of the most important integrable systems \cite{d,119}. It is
well-known soliton equations such as the KdV, modified KdV, and nonlinear Schr\"{o}dinger
equations are closely related to or derived from the Toda equation by
suitable limiting procedures \cite{119,120}. Various kinds of Toda lattice have
been discussed since it was proposed \cite{130,131,119,122,151}.
Among them, a remarkable discovery was made by Ruijsenaars in the area of
integrable lattice systems \cite{a}.
He found a relativistic integrable generalization of non-relativistic Toda lattice
through solving a relativistic
version of the Calogero-Moser system.
The Lax representation, inverse scattering problem of the Ruijsenaars-Toda
lattice and its
connection with soliton dynamics  were investigated.
A general approach to construct relativistic generalizations of integrable
lattice systems, applicable to the whole lattice KP hierarchy, was proposed by
Gibbons and Kupershmidt \cite{i}. After that, a series of techniques to
construct relativistic lattice equations were developed by systematic
procedure and Hirota's bilinear method \cite{c,l}.

The Ruijsenaars-Toda lattice, sometimes also called relativistic Toda (RT)
lattice, takes the form \cite{c}
\begin{equation}\label{1.1}
    \begin{split}
    &\beta_{t}=(1+\hbar\beta)(\alpha-\alpha^-),\\
    &\alpha_{t}=\alpha\left(\beta^+-\beta+\hbar\alpha^+-\hbar
    \alpha^-\right)\\
    \end{split}
\end{equation}
in Flaschka variables or
\begin{equation}\label{1.2}
\begin{split}
   & x_{k,tt}=\left(1+\h x_{k,t}\right)\left(1+\h x_{k+1,t}\right)
   \frac{e^{x_{k+1}-x_k}}{1+\h^2e^{x_{k+1}-x_k}}\\
   & -(1+\h x_{k-1,t})(1+\h x_{k,t})\frac{e^{x_k-x_{k-1}}}{1+\h^2e^{x_k-x_{k-1}}}
\end{split}
\end{equation}
in Newtonian form, where the small time step $\hbar=c^{-1}$ and $c$ is light speed.
In the non-relativistic limit $c\rightarrow\infty$, the RT equation (\ref{1.1}) reduced to
the well-known Toda lattice equation \cite{d},
\begin{equation}\label{1.3}
    \beta_{t}=\alpha-\alpha^-,~~\alpha_{t}=\alpha(\beta^+-\beta)
\end{equation}
in Flaschka variables, or
\begin{equation}\label{1.4}
    x_{k,tt}=e^{x_{k+1}-x_k}-e^{x_k-x_{k-1}}
\end{equation}
in Newtonian form. Eq (\ref{1.1}) is the Poincare-invariant generalizations of the
Galilei-invariant Toda systems (\ref{1.3}).

Mathematical frame work such as Lax representation, B\"{a}cklund transformation,
Hamiltonian structure of RT lattice
Eq (\ref{1.1}), etc, were investigated by some authors \cite{e}-\cite{k}.
Cosentino obtained a soliton solution by
using the IST method.  Hietarinta and Junkichi Satsuma transformed the RT
eq (\ref{1.1}) into trilinear form through
a suitable dependent variable transform. Later Yasuhiro Ohta, etc. decomposed
the RT lattice eq (\ref{1.1}) into three
Toda systems,  the Toda lattice itself, B\"{a}cklund transformation of Toda
lattice, and discrete time Toda lattice and
explicitly derived the solutions in terms of the Casorati determinant\cite{l}.
The solution they obtained converges
to that of TL eq (\ref{1.3}) in the limit of $c\rightarrow\infty$.

Algebro-geometric solutions (finite-gap solutions or
quasi-period solutions),  as  an important character of integrable system,
is  a kind of explicit solutions closely related to the inverse spectral
theory \cite{t,99}.
Around 1975, several independent groups in UUSR and USA, namely, Novikov,
Dubrovin and Krichever in Moscow, Matveev and Its in Leningrad, Lax, McKean,
van Moerbeke and
M. Kac in New York, and Marchenko, Kotlyarov and Kozel in Kharkov, developed
the so-called finite finite-gap theory of
nonlinear  KdV   equation based on
the works of Drach, Burchnall and Chaunchy, and Baker \cite{21,4,5}.
The algebro-geometric method they established allowed us to find an important
class of exact solutions to the soliton equations.
As a degenerated case of this solutions, the multisoliton solutions and
elliptic functions may be obtained \cite{t,1}.
Its and Matveev first derived explicit expression of the quasi-period
solution of KdV equation in 1975 \cite{4}, which is closely related to
the finite-gap spectrum of the associated differential operator. Further
exciting results appeared later, including the finite-gap solutions of
Toda lattice,  the Kadomtsev-Petviashvili  equation and others \cite{d,5,1},
which could be
found in the wonderful work of Belokolos, et al \cite{t}.  In recent years,
a systematic approach based on the nonlinearization technique
of Lax pairs or the restricted flow technique to derive the algebro-geometric
solutions of $(1+1)-$ and $(2+1)-$dimensional soliton equations has been
obtained
\cite{6}-\cite{11}. An alternate systematic approach proposed by Gesztesy
and Holden can be used to construct  algebro-geometric solutions has been
extended to the whole (1+1) dimensional continuous and discrete hierarchy
models  \cite{u}-\cite{n},\cite{13,19}.

In this paper, we mainly discussed the algebro-geometric quasi-period
solutions of RT hierarchy for fixed
constant $\h=c^{-1}\neq 0$.  In following section 2,
the RT equation (\ref{1.1}) is extended to a whole RT hierarchy through
the polynomial recursive relations.
In section 3,
we give a detailed study of algebro-geometric solutions for the
stationary RT hierarchy.
Firstly we derived the hyperelliptic curve  
$\mathcal{K}_p$ in connection
with the stationary RT hierarchy.
Then a fundamental meromorphic function $\phi$  on $\mathcal{K}_p$ ,
the Baker-Akhiezer
vector $\Psi$ and  the common eigenfunction of zero-curvature pair $U, V_p$,
was introduced to  study the trace formula and asymptotic properties of
$\phi$ and $\psi_1$, respectively. 
With the help of Riemann theta function
associated with $\mathcal{K}_p$, one finds the
theta function representations for $\phi$ and $\psi_1$ by alluding to Riemann's
vanishing theorem and the Riemann-Roch
theorem.  In section 4, we derive the complex-valued algebro-geometric
solutions of RT hierarchy with a given inital value problem by using the results in
sections 3 and 4. Finally, in Appendix A and B we give Lagrange interpolation
representation and asymptotic spectral parameter expansions 
that will be used in this paper.

\section{The Ruijsenaars-Toda Hierarchy }\label{section2}
In this section, we derive the Ruijsenaars-Toda hierarchy
by using a polynomial recursion formalism.
Throughout this section let us make
the following assumption.

\newtheorem{hyp1}{Hypothesis}[section]
\begin{hyp1}
\emph{In stationary case we assume that $u:\mathbb{R}\rightarrow\mathbb{C}$ satisfies
\begin{equation}\label{1}
\alpha,\beta\in\mathbb{C}^{\mathbb{Z}},~~\alpha(n)\neq 0,~~n\in\mathbb{Z}
\end{equation}
In the time-dependent case we suppose  $u:\mathbb{R}^2\rightarrow\mathbb{C}$
satisfies
\begin{eqnarray}
&&\alpha(\cdot,t),\beta(\cdot,t)\in\mathbb{C}^{\mathbb{Z}},\quad \alpha(n,t)\neq 0,\quad (n,t)
\in\mathbb{Z}\times\mathbb{R}\nonumber\\
&&\alpha(n,\cdot),\beta(n,\cdot)\in C^1(\mathbb{R}).
\end{eqnarray}
}
\end{hyp1}

In this paper, we denote by $S^{\pm}$ the shift operators acting on
$\psi=\{\psi(n)\}_{n=-\infty}^{+\infty}\in\mathbb{C}^{\mathbb{Z}}$ according
to $(S^{\pm}\psi)(n)=\psi(n\pm1),$ or $\psi^{\pm}=S^{\pm}\psi$
for convenience.

We start from the following $2\times2$ matrix isospectral problem \cite{p}
\begin{equation}\label{2.1}
    S^+\psi=U\psi,\quad \psi=\left(\begin{array}{c}
                              \psi_1 \\[0.2cm]
                              \psi_2
                              \end{array}\right),\quad
    U=\left(
        \begin{array}{ccc}
          0 & 1 \\[0.2cm]
          (\hbar z-1)\alpha & z+\beta \\
           \end{array}
      \right),
 \end{equation}
where the functions $\alpha, \beta$ are potentials, and $z$
is a constant spectral parameter independent of variable $n$.

Define sequences $\{f_\ell\}_{\ell\in\mathbb{N}_0}$ and $\{g_\ell\}_{\ell\in\mathbb{N}_0}$
recursively by
\eqnarrayl{2.2}
{&&g_{0}=-1/2,~f_{0}=0,\label{2.2a}\\
&& f_{\ell+1}+\beta f_\ell+g_\ell+g_\ell^-=0, ~~\ell\in\mathbb{N}_0,\label{2.2b}\\
&& \hbar\alpha f_{\ell+1}^{-}-\hbar\alpha^+f_{\ell+1}^+-\alpha f_\ell^-
+\alpha^+f_\ell^+-\beta g_\ell^-+\beta g_\ell\nonumber \\
&&-g_{\ell+1}^-+g_{\ell+1}=0, ~~ \ell\in\mathbb{N}_0\label{2.2d}.
}
Explicitly, one obtains
\eqnarrayl{2.3}
{f_{1}&=&1,~~g_{1}~=~\hbar\alpha^{+}-\delta_{1}/{2},\nonumber\\
f_{2}&=&-\hbar(\alpha^++\alpha)-\beta+\delta_1,\nonumber\\
g_{2}&=&-\hbar^2\alpha^{++}\alpha^+-\hbar^2(\alpha^+)^2+\hbar^2\alpha^
+\alpha-\hbar\alpha^+\beta^+-\hbar\alpha^+\beta-\alpha^+\nonumber\\
&+&\hbar\alpha^+\delta_1-\delta_2/2,\nonumber\\
f_{3}&=&\hbar^2\alpha^{++}\alpha^{+}+\hbar^2(\alpha^+)^2
-\hbar^2\alpha^+\alpha+\hbar\alpha^+\beta^++2\hbar\alpha^+\beta+\alpha^+\\
&+&\hbar^2\alpha^+\alpha+\hbar^2\alpha^2-\hbar^2\alpha\alpha^-
+2\hbar\alpha\beta+\hbar\alpha\beta^-+\alpha-(\hbar\alpha^+\nonumber\\
&+&\hbar\alpha+\beta)\delta_1+\delta_2,~~\text{etc.}\nonumber
 }
Here $\{\delta_\ell\}_{\ell\in\mathbb{N}}$ denote summation constants
which naturally arise when solving (\ref{2.2}). Subsequently, it will
be useful to work with the corresponding homogeneous coefficients
$\hat{f}_\ell, \hat{g}_\ell$ defined by the vanishing of all
summation constants $\delta_k$ for $k=1,\ldots\ldots, \ell,$
\eqnarrayn{
&&\hat{g}_{0}=-1/2,~~\hat{g}_{\ell}=g_\ell|_{\delta_{j}=0,j=1,\dotsi,\ell},~~ \ell\in\mathbb{N},\\
&&\hat{f}_0=0,~~\hat{f}_{\ell}=f_\ell|_{\delta_{j}=0,j=1,\dotsi,\ell},~~ \ell\in\mathbb{N}.}
By induction one infers that
\eqnarrayl{2.3a}
{\begin{split}
&g_{\ell}=\sum_{s=0}^{\ell}\delta_{\ell-k}\hat{g}_{k},~~\ell\in\mathbb{N}_{0}\\
&f_{\ell}=\sum_{s=0}^{\ell}\delta_{\ell-k}\hat{f}_{k},~~\ell\in\mathbb{N}_{0},
\end{split}}
introducing
\eqn{\delta_0=1.}

\newtheorem{rem1}[hyp1]{Remark}
\begin{rem1}\emph{
The constants $\delta_j~(j\in\mathbb{N}_0),$ can be expressed
in terms of the branch points $E_i,~(i=0,\dotsi,p+1)$ of the associated
spectral curve defined in $(\ref{3.5})$. $($Theorem $6)$}
\end{rem1}

In order to obtain the RT hierarchy associated with the spectral
problem (\ref{2.1}), we first solve the stationary zero-curvature equation
\eql{2.4}{U V_p- V^{+}_p U=0,~~V_p=(V_{ij})_{2\times 2}}
with
\eql{2.5}
{V_p=\left(
       \begin{array}{cc}
         V_{11}^- & V_{12}^- \\[0.2cm]
         V^-_{21} & V^-_{22} \\
       \end{array}
     \right),
}
where each entry $V_{ij}$ is a polynomial in $z,$
\eqnarrayl{2.6}
{
V_{11} &=&\sum_{j=0}^{p+1}g_{p+1-j} z^j+
f_{p+2},
~~V_{12} ~=~\sum_{j=0}^{p+1}f_j z^{p+1-j} ,\label{2.6a}\\
V_{21} &=&(\hbar z-1)\alpha^+V_{12}^+
~=~(\hbar z-1)\alpha^+\left(\sum_{j=0}^{p+1}f_j^+z^{p+1-j}\right),
 \label{2.6b}\\
V_{22} &=&-\sum_{j=0}^{p+1}g_{p+1-j}z^j.\label{2.6c}
}
Equation (\ref{2.4}) can be rewritten as
\eqnarrayl{2.7}{
&&V_{21}^--(\hbar z-1)\alpha V_{12}=0,\label{2.7a}\\
&&V_{22}^--V_{11}-(z+\beta)V_{12}=0,\label{2.7b}\\
&&(\hbar z-1)\alpha V_{11}^-+(z+\beta)V_{21}^--(\hbar z-1)\alpha V_{22}=0,\label{2.7c}\\
&&(\hbar z-1)\alpha V_{12}^-+(z+\beta)V_{22}^--V_{21}-(z+\beta)V_{22}=0.\label{2.7d}
}
Since det($U$)$\neq 0$ for $z\in\mathbb{C}\backslash\{1/\hbar\}$ by (\ref{1}),
(\ref{2.4}) yields $\text{tr}(V_p^+)=\text{tr}(UV_pU^{-1})=\text{tr}(V_p)$
and hence
\eqn{V_{11}^-+V_{22}^-=V_{11}+V_{22},}
implying $V_{11}+V_{22}$ is a lattice constant.
If $V_{11}+V_{22}=c\neq 0$, one can add a polynomial times
the identity to $V_p,$ which does not affect the zero-curvature
equation. Therefore we can choose \eql{2.4k}{V_{11}+V_{22}=0}
without loss of generality.
This fact leads to the following result.



\newtheorem{lem2}[hyp1]{Lemma}
\begin{lem2}\label{lemm2}\emph{
Suppose $U$ and $V_p$ satisfy the stationary zero-curvature equation $(\ref{2.4})$.
Then $(\ref{2.7a})$-$(\ref{2.7d})$ change into
\eqnarrayl{2.14}{
&&  V_{11}^-+  V_{11}+(z+\beta)  V_{12}=0,\label{2.14a}\\
&&(\hbar z-1)\alpha V_{12}^--(z+\beta)V_{11}^--(\hbar z-1)\alpha^+V_{12}^+
+(z+\beta)V_{11}=0.\nonumber\\\label{2.14b}
 }
In particular,
the coefficients $\{f_\ell\}_{\ell=0,\ldots,p}$ and $\{g_\ell\}_{\ell=0,\ldots,p}$ defined in
$(\ref{2.6a})$-$(\ref{2.6c})$
satisfy the recursive relations $(\ref{2.2a})$-$(\ref{2.2d}).$
}
\end{lem2}
\noindent{\bf Proof.} Eq.(\ref{2.14a}) and eq.(\ref{2.14b}) arise from (\ref{2.7a})-(\ref{2.7d}) by
substituting $V_{22}$ for $-V_{11}.$ Insertion of (\ref{2.6a})-(\ref{2.6c}) into (\ref{2.14a}) and
(\ref{2.14b}) then yields the relations $(\ref{2.2a})$-$(\ref{2.2d}).$ \qed

Inserting (\ref{2.6a})-(\ref{2.6c}) into (\ref{2.4})
and using the results of Lemma \ref{lemm2} yield
the following theorem.

\newtheorem{them1}[hyp1]{Theorem}
\begin{them1}\emph{
Suppose that $U$ and $V_p$ satisfy the stationary zero-curvature equation $(\ref{2.4})$.
Then
$(\ref{2.4})$ reads
\eqnarrayl{2.8}{
0&=&UV_p-V^+_pU\nonumber\\
&=&\left(
     \begin{array}{cccc}
       0 & 0 \\ [0.4cm]
       \begin{array}{c}
          (\hbar z-1)(\alpha f_{p+2}^- \\
          -\alpha f_{p+2}) \\
         \end{array}
        & \begin{array}{cc}-\alpha f_{p+1}^-+\alpha^+f_{p+1}^+\\
        -\beta(g_{p+1}^--g_{p+1})\end{array}\\
     \end{array}
   \right),
}
which is equivalent to
\eqnarrayl{2.9}
{&&f_{p+2}^--f_{p+2}=0,\label{2.9a}\\
&&\beta(g_{p+1}^--g_{p+1})+\alpha f_{p+1}^--\alpha^+f_{p+1}^+=0.\label{2.9b}}
}
\end{them1}

Thus, varying $p\in\mathbb{N}_0,$ equations (\ref{2.9a}) and (\ref{2.9b})
give rise to the stationary Ruijsenaars-Toda (RT) hierarchy which we introduce
as follows
\eql{2.12}{\text{s-RT}_p(\alpha,\beta)
=\left(
   \begin{array}{c}
     f_{p+2}^--f_{p+2} \\[0.2cm]
     \beta(g_{p+1}^--g_{p+1})+\alpha f_{p+1}^--\alpha^+f_{p+1}^+ \\
   \end{array}
 \right)=0,~~p\in\mathbb{N}_0.
}
In the special case $p=0,$ one obtains the stationary
version of the Ruijsenaars-Toda system (\ref{1.1})
\eqnarrayl{2.13}{
\left(
\begin{array}{cc}
\hbar\alpha(\alpha^--\alpha^+)+\alpha(\beta^--\beta)\\[0.2cm]
\hbar\beta(\alpha-\alpha^+)-\alpha+\alpha^+\\
\end{array}
\right)=0.
}
In the case $p=1,$
one finds
\eqnarrayn{&&
\left(
 \begin{array}{cc}\begin{array}{cc}
\alpha(\hbar^2\alpha^{++}\alpha^++\hbar^2(\alpha^+)^2-\hbar^2\alpha^+\alpha+\hbar\alpha^+\beta^++\alpha^+\\
+\hbar\alpha\beta-\hbar\alpha\beta^--\hbar^2\alpha\alpha^--\hbar^2(\alpha^-)^2+\hbar^2\alpha^-\alpha^{--}\\
-2\hbar\alpha^-\beta^--\hbar\alpha^-\beta^{--}-\alpha^-)\\
 \end{array}\\[1cm]\begin{array}{cc}
    -\hbar\alpha^2-\hbar\alpha\alpha^--\alpha\beta^-+\hbar\alpha^+\alpha^{++}+\hbar(\alpha^+)^2\\
    +\alpha^+\beta^++\beta(-\hbar^2\alpha^+\alpha-\hbar^2\alpha^2+\hbar^2\alpha\alpha^-\\
    -\hbar\alpha\beta-\hbar\alpha\beta^--\alpha+\hbar^2\alpha^{++}\alpha^{+}+\hbar^2(\alpha^+)^2\\
    -\hbar^2\alpha^+\alpha+\hbar\alpha^+\beta^++\hbar\alpha^+\beta+\alpha^+)\\
    \end{array}
  \end{array}
\right)\\
&&+\delta_1\left(
     \begin{array}{cc}
       \alpha(-\hbar\alpha^+-\beta+
      \hbar\alpha^-+\beta^-) \\[0.2cm]
      \alpha-\alpha^+
     +\hbar\alpha\beta-\hbar\alpha^+\beta  \\
     \end{array}
   \right)
 =0.
}
In accordance with our notation introduced in (\ref{2.3a}),
the corresponding homogeneous stationary Ruijsenaars-Toda equations
are defined by
\eqn{\textrm{s-$\widehat{\text{RT}}_p(\alpha,\beta)$}
=\textrm{$\text{s-RT}_p(\alpha,\beta)$}|_{\delta_0=1,\delta_\ell=0,
\ell=1,\ldots,p},~~p\in\mathbb{N}_0.}

Next we turn to the time-dependent Ruijsenaars-Toda hierarchy. For that purpose
the potentials $\alpha$ and $\beta$ are now considered as functions of both the
lattice point and time. For each equation in the hierarchy, that is, for each
$p\in\mathbb{N}_0,$ we introduce a deformation (time) parameter $t_p\in\mathbb{R}$
in $\alpha,\beta,$ replacing $\alpha(n),$$\beta(n)$ by $\alpha(n,t_p),$$\beta(n,t_p).$
The quantities $\{f_\ell\}_{\ell\in\mathbb{N}_0}$ and $\{g_\ell\}_{\ell\in\mathbb{N}_0}$
are still defined by $(\ref{2.2a})$-$(\ref{2.2d}).$
The time-dependent Ruijsenaars-Toda hierarchy then obtained by
imposing the zero-curvature equations
\eql{2.15}{U_{t_p}(t_p)+U(t_p)V_p(t_p)-V_p^+(t_p)U(t_p)=0,~~t_p\in\mathbb{R}.}
Relation (\ref{2.15}) implies
\eqnarrayl{2.16}{
\left(
\begin{array}{cc}
0&0\\[0.2cm]
(\hbar z-1)\alpha_{t_{p}}&\beta_{t_p}\\
\end{array}
\right)+
\left(
\begin{array}{cc}
0&0\\[0.2cm]
\begin{smallmatrix}
(\hbar z-1)(\alpha f_{p+2}^-\\
-\alpha f_{p+2})
\end{smallmatrix}
&\begin{smallmatrix}
-\alpha f_{p+1}^-
+\alpha^+f_{p+1}^+\\
-\beta(g_{p+1}^--g_{p+1})
\end{smallmatrix}
\end{array}
\right)=0.
}
Varying $p\in\mathbb{N}_0,$ the collection of
evolution equations
\eqnarrayl{2.17}{
&&\text{RT}_p(\alpha,\beta)=\left(
 \begin{array}{c}
 \alpha_{t_p}-\alpha(f_{p+2}+f_{p+2}^-) \\[0.2cm]
 \beta_{t_p}-\beta(g_{p+1}^--g_{p+1})-\alpha f_{p+1}^-+\alpha^+ f_{p+1}^+ \\
  \end{array}
  \right)=0,\nonumber\\
  &&~~~~~~~~~~~~~~~~~~~~~~~~~~~~~~~~~~~~~~~~~~~~
  (n,t_p)\in\mathbb{Z}\times\mathbb{R},~~p\in\mathbb{N}_0,
}
then defines the time-dependent Ruijsenaars-Toda hierarchy.
Explicitly,
\eqnarrayn{\text{RT}_0(\alpha,\beta)&=&
\left(
\begin{array}{cc}
\alpha_{t_0}-\hbar\alpha(\alpha^--\alpha^+)-\alpha(\beta^--\beta)\\[0.2cm]
\beta_{t_0}-\hbar\beta(\alpha-\alpha^+)+\alpha-\alpha^+\\
\end{array}
\right)=0,\\[0.2cm]
\text{RT}_1(\alpha,\beta)&=&
\left(
  \begin{array}{cc}\begin{array}{cc}
  \alpha_{t_1}-\alpha(\hbar^2\alpha^{++}\alpha^++\hbar^2(\alpha^+)^2-\hbar^2\alpha^+\alpha\\
  +\hbar\alpha^+\beta^++\alpha^++\hbar\alpha\beta-\hbar\alpha\beta^--\hbar^2\alpha\alpha^-\\
  -\hbar^2(\alpha^-)^2+\hbar^2\alpha^-\alpha^{--}-2\hbar\alpha^-\beta^-\\-\hbar\alpha^-\beta^{--}
  -\alpha^-)\\
  \end{array}\\[1cm]\begin{array}{cc}
    \beta_{t_1}+\hbar\alpha^2+\hbar\alpha\alpha^-+\alpha\beta^--\hbar\alpha^+\alpha^{++}-\hbar(\alpha^+)^2\\
    -\alpha^+\beta^+-\beta(-\hbar^2\alpha^+\alpha-\hbar^2\alpha^2+\hbar^2\alpha\alpha^-\\
    -\hbar\alpha\beta-\hbar\alpha\beta^--\alpha+\hbar^2\alpha^{++}\alpha^{+}+\hbar^2(\alpha^+)^2\\
    -\hbar^2\alpha^+\alpha+\hbar\alpha^+\beta^++\hbar\alpha^+\beta+\alpha^+)\\\end{array}
  \end{array}
\right)\\
&+&\delta_1\left(
  \begin{array}{c}
    -\alpha(-\hbar\alpha^+-\beta+
  \hbar\alpha^-+\beta^-)\\[0.4cm]
    -(\alpha-\alpha^+
    +\hbar\alpha\beta-\hbar\alpha^+\beta) \\
  \end{array}
\right)
=0,~\text{etc.},}
represent the first two equations of the time-dependent
Ruijsenaars-Toda hierarchy. The system of equations, $\text{RT}_0(\alpha,\beta)=0$
is of course the Ruijsenaars-Toda system (\ref{1.1}).

Next, taking into account (\ref{2.4k}), one infers that
the expression $R_{2p+2},$ defined as
\eqnarrayl{2.18}{R_{2p+2}(z)&=&-V_{11}^{2}(z,n)-V_{12}(z,n)V_{21}(z,n),\nonumber\\
&=&-V_{11}^{2}(z,n)-(\hbar z-1)\alpha^+V_{12}(z,n)V_{12}^+(z,n)
}
is a lattice constant, that is, $R_{2p+2}-R_{2p+2}^-=0,$
since taking determinants in the stationary zero-curvature
equation (\ref{2.4}) immediately yields
\eqn{(\hbar z-1)\alpha\left((-V_{11}^-)^2-V_{12}^-
V_{21}^-+V_{11}^{2}+V_{12}V_{21}\right)=0.}
Hence, $R_{2p+2}(z)$
only depends on $z,$ and one may write
$R_{2p+2}$ as
\eql{3.5}{R_{2p+2}(z)=-\frac{1}{4}\prod_{m=0}^{2p+1}(z-E_m),
~~\{E_m\}_{m=0}^{2p+1}\in\mathbb{C},~~p\in\mathbb{N}_0.}
Relations (\ref{2.18}) and (\ref{3.5}) allows one to introduce
a hyperelliptic curve $\mathcal{K}_p$ of (arithmetic) genus
$p$ (possibly with a singular affine part), where
\eql{2.19}{\mathcal{K}_p:~~\mathcal{F}_p(z,y)
=y^2+4R_{2p+2}=y^2-\prod_{m=0}^{2p+2}(z-E_m)=0,~~p\in\mathbb{N}_0.}

Equations (\ref{2.14a}), (\ref{2.14b}) and (\ref{2.18}) permit one
to derive nonlinear difference equations for $V_{11}, V_{12}$ separately.
One obtains
\eqnarrayl{2.20}{
&&-V_{11}^2-(\hbar z-1)\alpha^+(V_{11}^++V_{11})(V_{11}+V_{11}^-)\nonumber\\
&&=(z+\beta)(z+\beta^+)R_{2p+2},\label{2.20a}\\
&&-\Big[(\hbar z-1)(\alpha V_{12}^--\alpha^+V_{12}^+)+(z+\beta)^2V_{12}\Big]^2\nonumber\\
&&-4(\hbar z-1)(z+\beta)^2
\alpha^+V_{12}V_{12}^+
=4(z+\beta)^2R_{2p+2}\label{2.20b}
}

Equations analogous to (\ref{2.20a}) and (\ref{2.20b})
can be used to derive nonlinear recursion relations
for homogenous coefficients $\hat{f}_\ell, \hat{g}_{\ell}$.
In addition, as proven in
Theorem B.1, (\ref{2.20a}) leads to an explicit
determination of the summation constants $\delta_1,\delta_2,\ldots,\delta_p$
in (\ref{2.12}) in terms of the zeros $E_0,\ldots,E_{2p+1}$
of associated polynomial $R_{2p+2}$ in (\ref{3.5}).
In fact, one can prove (cf.(\ref{ap15}))
\eql{2.21}{\delta_\ell=c_\ell(\underline{E}),~~\ell=0,\ldots,p,}
where
\eqnarrayn{
&&{c}_{0}(\underline{E})=1, ~~{c}_{1}(\underline{E})=-\frac{1}{2}\sum_{m=0}^{2p+1}E_{m}\\
&&c_k(\underline{E})=
\sum_{\begin{smallmatrix}j_0,\ldots,j_{2p+1}=0,\\j_0+\ldots+ j_{2p+1}=k\end{smallmatrix}
}^{k}\frac{(2j_0)!\ldots(2j_{2p+1})! E_0^{j_0}\dotsi E_{2p+1}^{j_{2p+1}}}{2^{2k}(j_0!)^2
\ldots (j_{2p+1}!)^2(2j_0-1)\dotsi (2j_{2p+1}-1)},\\
&&\quad~~~~~~~~~~~~~~~~~~~~~~~~~~~~~~~~~~~~~~~~~~~~~~~~~~~~~~~~~~~~~~~~~~~~~~~~~~
 k\in\mathbb{N}.
}
are symmetric functions of $\underline{E}=(E_0,\ldots,E_{2p+1}).$

\newtheorem{rem2.5}[hyp1]{Remark}
\begin{rem2.5}
\emph{If $\alpha, \beta$ satisfy one of the stationary Ruijsenaars-Toda equations in $(\ref{2.12})$
for a particular value of $p$, s-AL$_p(\alpha,\beta)=0$, then they satisfy infinitely
many such equations of order higher than $p$ for certain choices of summation constants
$\delta_\ell$. This is seen as follows.
Assume $f_{p+2}-f_{p+2}^-=0$ for some $p\in\mathbb{N}$ and some set of
integration constants $\{\delta_\ell\}_{\ell=1,\ldots,p}\subset\mathbb{C},$ one infers
\eqn{f_{p+2}=\lambda_{p+2}}
for some constant $\lambda_{p+2}\in\mathbb{C}.$ Subtracting the constant
$\lambda_{p+2}$ $($i.e.writing $f_{p+2}=\sum_{s=0}^{p+2}\breve{\delta}_{p+2-k}\hat{f}_{k}$,
for some set of constants $\{\breve{\delta}_\ell\}_{\ell=1,\ldots,p+2}$ and absorbing
$\lambda_{p+2}$ into $\breve{\delta}_{p+2}$$),$ we may without loss of generality
assume that $f_{p+2}=0,$ and
hence the recursion (\ref{2.2d}) implies
\eqn{\beta(g_{p+1}^--g_{p+1})+\alpha f_{p+1}^--\alpha^+f_{p+1}^+=0} in (\ref{2.12})
is equivalent to
\eqn{g_{p+2}-g_{p+2}^--\hbar \alpha^+f_{p+2}^--\hbar \alpha f_{p+2}^-=0.}
Hence, $$g_{p+2}=\lambda^{*}_{p+2}$$ for some constant $\lambda^{*}_{p+2}\in\mathbb{C}.$
This indicates
\eqn{\beta(g_{p+2}^--g_{p+2})+\alpha f_{p+2}^--\alpha^+f_{p+2}^+=0.}
Then \eqn{f_{p+3}=\lambda_{p+3}} for some constant $\lambda_{p+3}\in\mathbb{C}$ which
arises from (\ref{2.2b}).
Similarly,
subtracting the constant
$\lambda_{p+3}$ (i.e.writing $f_{p+3}=\sum_{s=0}^{p+3}\tilde{\delta}_{p+3-k}\hat{f}_{k}$,
for some set of constants $\{\tilde{\delta}_\ell\}_{\ell=1,\ldots,p+3}$ and absorbing
$\lambda_{p+3}$ into $\tilde{\delta}_{p+3}$), we may without loss of generality
assume that $f_{p+3}=0,\ldots$ Iterating this procedure yields
\eqn{\text{s-RT}_q(\alpha,\beta)=0} for all $q\geq p+1 $ (corresponding to some $p$-dependent
choice of integration constants $\{\breve{\delta}_\ell\}_{\ell=1,\ldots,p}$).
}
\end{rem2.5}

\section{Stationary Algebro-geometric Solutions}\label{section3}
This section is devoted to a detailed study of the stationary Ruijsenaars-Toda hierarchy and
its algebro-geometric solutions. Our basic tools are derived from combining
the polynomial recursion formalism introduced in Section \ref{section2} and a fundamental
meromorphic function $\phi$ on a hyperelliptic curve $\mathcal{K}_p.$ We
will obtain explicit Riemann theta function representations for the
meromorphic function $\phi$,
the Baker-Akhiezer function $\psi_1$, and the algebro-geometric
solutions $\alpha,\beta$.

Unless explicitly stated otherwise, we suppose in this section that
\eql{3.1}{\alpha,\beta\in\mathbb{C}^{\mathbb{Z}},~~\alpha(n)\neq 0,~~n\in\mathbb{Z}}
and assume (\ref{2.2a})-(\ref{2.2d}), (\ref{2.4})-(\ref{2.6}), (\ref{2.14a})-(\ref{2.12}),
(\ref{2.18})-(\ref{2.19}), keeping $p\in\mathbb{N}_0$ fixed.

Throughout this section we assume $\mathcal{K}_p$ defined in (\ref{2.19})
to be nonsingular, that is, we suppose
that
\eql{3.2}{E_{m}\neq E_{m^{\prime}}~~\text{for}~~ m\neq m^{\prime},~~ E_m\in\mathbb{C}\backslash\{\h\},
~~m=0,\ldots,2p+1.}
We compactify
$\mathcal{K}_p$ by adding two points $P_{\infty+}$ and $P_{\infty-}$,
$P_{\infty+}\neq P_{\infty-}$, at infinity, still denoting its
projective closure by $\mathcal{K}_p$.
Finite points $P$ on $\mathcal{K}_p$ are denoted by $P=(z, y)$ where
$y(P)$ denotes the meromorphic function on $\mathcal{K}_p$ satisfying $\mathcal{F}_p(z, y)=0$.
The complex
structure on $\mathcal{K}_p$ is then defined in a standard manner and $\mathcal{K}_p$
has topological genus
$p$. Moreover, we use the involution
\eql{3.3}{*: \mathcal{K}_p\rightarrow\mathcal{K}_p, ~~P=(z,y)\mapsto P^*=(z,-y),~~
 P_{\infty\pm}\mapsto P_{\infty\pm}^*=P_{\infty\mp}.}

 We also emphasize that by fixing the curve $\mathcal{K}_{p}$ (i.e., by fixing $E_0,
 \ldots,E_{2p+1}$), the summation constants \textrm{$\{\delta_\ell\}_{\ell=0,\ldots,p}$}
 in the stationary $\text{RT}_p$ equations are uniquely determined as is clear from
 (\ref{2.21}), which establish the summation constants $\delta_\ell$ as symmetric
 functions of $E_0,
 \ldots,E_{2p+1}.$

 For notational simplicity we will usually tacitly assume that $p\in\mathbb{N}.$(The
 trivial case $p=0$ is explicitly treated in Example 3.6)

 In the following, the zeros of the polynomial $V_{12}(\cdot,n)$ (cf. (\ref{2.6a}))
 will play a special role.
 We denote them by $\{\mu_j(n)\}_{j=1,\ldots,p}$ and hence
 write
 \eql{3.4}{V_{12}(z)=\prod_{j=1}^p(z-\mu_j).}
 Similarly we write
 \eqnarrayl{3.5a}{V_{21}(z)=\hbar\alpha^+(z-\frac{1}{\hbar})\prod_{j=1}^{p}(z-\mu_{j}^{+}),~~
 \mu_{j}^{+}(n)=\mu_{j}(n+1),\\
 ~~~~~~~~~~~~~~~~~~j=1,\ldots,p,~~ n\in\mathbb{Z},\nonumber
 }
 and we recall that (cf.(\ref{2.18}))
 \eql{3.6}{R_{2p+2}+V_{11}^2=-(\hbar z-1)\alpha V_{12}V_{12}^+.}
 The next step is crucial; it permits us to "lift"
 the zeros $\mu_j$ from the complex plane $\mathbb{C}$ to the
 curve $\mathcal{K}_p$. From (\ref{3.6}) one infers
 that
 \eqn{R_{2p+2}(z)+V_{11}(z)^2=0,~~z\in\{\mu_j,\mu_j^+,1/\hbar\}_{j=1,\ldots,p}.}
Now we introduce $\{\mu_j\}_{j=1,\ldots,p}\subset\mathcal{K}_p,$
$\{\mu_j^+\}_{j=1,\ldots,p}\subset\mathcal{K}_p$ and $P_{\hbar}\in\mathcal{K}_p$
by
\begin{flalign}
&\hat{\mu}_{j}(n)=(\mu_{j}(n),-2V_{11}(\mu_{j}(n),n)),~~j=1,\ldots,p~~n\in\mathbb{Z}, \label{3.7a}\\
&\hat{\mu}_{j}^{+}(n)=(\mu_{j}^{+}(n),2V_{11}(\mu_{j}^+(n),n)),~~j=1,\ldots,p~~n\in\mathbb{Z},\label{3.7b}
\end{flalign}
and
\eql{3.8}{P_{\hbar}=(1/\hbar,2V_{11}(1/\hbar,n)),}
where
$$(2V_{11}(1/\hbar,n))^2=y(1/\hbar)^2=\prod_{m=0}^{2p+1}(\hbar-E_m)$$
is independent on $n.$

Next we briefly define some notations in connection
with divisors on $\mathcal{K}_p$ \cite{m,n}.
A map, $\mathcal{D}:\mathcal{K}_p\rightarrow\mathbb{Z},$ is called a divisor
on $\mathcal{K}_p$ if $\mathcal{D}(P)\neq 0$ for only finitely
many $P\in\mathcal{K}_p.$ The set of divisors on $\mathcal{K}_p$
is denoted by $\text{Div}(\cur).$ We shall employ the following
(additive) notation for divisors,
\eqnarrayn{
&&\mathcal{D}_{Q_0\underline{Q}}=\mathcal{D}_{Q_0}+\mathcal{D}_{\underline{Q}},~~
\mathcal{D}_{\underline{Q}}=\mathcal{D}_{Q_1}+\ldots+\mathcal{D}_{Q_m},\\
&&\underline{Q}=\{Q_1,\ldots,Q_m\}\in\text{Sym}^m\cur,~~Q_0\in\cur,~m\in\mathbb{N},
}
where for any $Q\in\cur,$
\eqn{\mathcal{D}_{Q}:\cur\rightarrow\mathbb{N}_0,~~
P\mapsto \mathcal{D}_Q(P)=
\begin{cases}
1&\text{for}~~P=Q\cr
0&\text{for}~~P\in\cur\backslash\{ Q\},
\end{cases}
}
and $\text{Sym}^n\cur$ denotes the $n$th symmetric product of $\cur.$
In particular, one can identify $\text{Sym}^m\cur$ with the set of
nonnegative divisors $0\leq\mathcal{D}\in\text{Div}(\cur)$ of degree
$m$. Moreover, for a nonzero, meromorphic function
$f$ on $\cur$, the divisor of $f$ is denoted by $(f).$
Two divisors $\mathcal{D},\mathcal{E}\in\text{Div}(\cur)$ are called
equivalent, denoted by $\mathcal{D}\thicksim\mathcal{E},$ if and only
if $\mathcal{D}-\mathcal{E}=(f)$ for some $f\in\mathcal{M}(\cur)\backslash\{0\}.$
The divisor class $[\mathcal{D}]$ of $\mathcal{D}$ is
then given by $[\mathcal{D}]=\{\mathcal{E}\in\text{Div}(\cur)
|\mathcal{D}\thicksim\mathcal{E}\}.$ We recall that
\eqn{\text{deg}((f))=0,~~f\in\mathcal{M}(\cur)\backslash\{0\},}
where the degree deg($\mathcal{D}$) of $\mathcal{D}$ is given
by $\text{deg}(\mathcal{D})=\sum_{P\in\cur}\mathcal{D}(P).$

Next we introduce the fundamental meromorphic function on $\mathcal{K}_p$ by
\eqnarrayl{3.9}{
\phi(P,n)&=&\frac{ y/2-V_{11}(z,n)}{V_{12}(z,n)}\label{3.9a}\\
&=&\frac{(\hbar z-1)\alpha^+ V_{12}^+(z,n)}{y/2+V_{11}(z,n)},
~~P=(z,y)\in\mathcal{K}_{p},~ n\in\mathbb{Z},\label{3.9b}}
with divisor $(\phi(\cdot,n))$ of $\phi(\cdot,n)$ given
by
\eql{3.10}{(\phi(\cdot,n))=\mathcal{D}_{P_{\hbar}\underline{\hat{\mu}}^{+}(n)}
-\mathcal{D}_{P_{\infty-}\underline{\hat{\mu}}(n)},}
using (\ref{3.4}) and (\ref{3.5a}).
Here we abbreviated
\eqn{\hat{\underline{\mu}}=\{\hat{\mu}_1,\ldots,\hat{\mu}_p\},
~~\hat{\underline{\mu}}^+=\{\hat{\mu}^+_1,\ldots,\hat{\mu}_p^+\}.}
Given $\phi(\cdot,n),$ the meromorphic stationary Baker-Akhiezer
vector
$\Psi(\cdot,n,n_0)$ on $\cur$ is then defined by
\begin{flalign}
\Psi(P,n,n_{0})=&
\left(\begin{array}{c}\psi_{1}(P,n,n_{0})\\\psi_{2}(P,n,n_{0})\end{array}\right),\label{3.11a}\\
\psi_{1}(P,n,n_{0})=&
\begin{cases}
\prod_{n^{\prime}=n_{0}}^{n-1}\phi(P,n^{\prime}),&n^{\prime}>n_{0},\cr
1,&n^{\prime}=n_{0},\cr
\prod_{n^{\prime}=n}^{n_{0}-1}\phi(P,n^{\prime})^{-1},&n^{\prime}<n_{0},
\end{cases}\label{3.11b}\\
\psi_{2}(P,n,n_{0})=&\phi(P,n_{0})\times
\begin{cases}
\prod_{n^{\prime}=n_{0}+1}^{n-1}\left(\frac{\alpha(n^{\prime})(\hbar z-1)}{\phi^{-}(P,n^{\prime})}
+z+\beta(n^{\prime})\right),&n^{\prime}>n_{0},\cr
1,&n^{\prime}=n_{0},\cr
\prod_{n^{\prime}=n+1}^{n_{0}}\left(\frac{\alpha(n^{\prime})(\hbar z-1)}{\phi^{-}(P,n^{\prime})}+z
+\beta(n^{\prime})\right)^{-1},&n^{\prime}<n_{0}.
\end{cases}\label{3.11c}
\end{flalign}
Basic properties of $\phi$ and $\Psi$ are summarized in
the following result.

\newtheorem{lem3.1}{Lemma}[section]
\begin{lem3.1}\label{lemma3.1}
\emph{
  Suppose $\al,\be$ satisfy (\ref{3.1}) and the $p$th
  stationary Ruijsenaars-Toda system (\ref{2.12}).
  Moreover, assume (\ref{3.5}) (\ref{2.19}) and (\ref{3.2}) and
  let $P=(z,y)\in\cur\backslash\{P_{\infty\pm},P_\h\},(n,n_0)\in\mathbb{Z}^2.$
  Then $\phi$ satisfies the Riccati-type equation
  \eql{3.12}{\phi(P)\phi^{-}(P)-(z+\beta)\phi^{-}(P)-(\h z-1)\alpha=0,}
  as well as
  \eqnarrayl{3.13}{
   &&\phi(P)\phi(P^{*})=-\frac{(\h z-1)\alpha^+ V_{12}^+(z)}{V_{12}(z)},\label{3.13a}\\
   &&\phi(P)+\phi(P^{*})=-2\frac{V_{11}(z)}{V_{12}(z)},\label{3.13b}\\
    &&\phi(P)-\phi(P^{*})=\frac{y}{V_{12}(z)}.\label{3.13c}}
    The vector $\Psi$ satisfies
    \begin{flalign}
   & \psi_{2}(P,n,n_{0})=\psi_{1}(P,n,n_{0})\phi(P,n),\label{3.14a}\\
    & U(z)\Psi^{-}(P)=\Psi(P),\label{3.14b}\\
    &V_p(z)\Psi^{-}(P)=(y/2)\Psi^-(P),\label{3.14c}\\
    &\psi_1(P,n,n_0)\psi_1(P^*,n,n_0)=(\h z-1)^{n-n_0}\Gamma(\al^+,n,n_0)\nonumber\\
    & ~~~~~~~~~~~~~~~~~~~~~~~~~~~~~~~~~~
    \times \frac{V_{12}(z,n)}{V_{12}(z,n_0)},\label{3.14d}\\
    &\psi_1(P,n,n_0)\psi_2(P^*,n,n_0)+\psi_1(P^*,n,n_0)\psi_2(P,n,n_0)\nonumber\\
    &~~~~~~~~~~~~~~~~~~~~~~~~~=-2(\h z-1)^{n-n_0}\Gamma(\al^+,n,n_0)
    \frac{V_{11}(z,n)}{V_{12}(z,n_0)},\label{3.14e}\\
    &\psi_1(P,n,n_0)\psi_2(P^*,n,n_0)-\psi_1(P^*,n,n_0)\psi_2(P,n,n_0)\nonumber\\
    &~~~~~~~~~~~~~~~~~~~~~~~~~=-(\h z-1)^{n-n_0}\Gamma(\al^+,n,n_0)\frac{y}{V_{12}(z,n_0)},\label{3.14f}
    \end{flalign}
    where we used the abbreviation
    \eql{3.15}{\Gamma(f,n,n_0)=
    \begin{cases}
    \prod_{n^\prime=n_0}^{n-1}f(n^\prime),&n> n_0,\cr
    1,&n=n_0,\cr
    \prod_{n^\prime=n}^{n_0-1}f(n^\prime)^{-1},&n< n_0.
    \end{cases}}
    }
\end{lem3.1}
\noindent {\bf Proof.} To prove (\ref{3.12}) one uses
the definitions (\ref{3.9a}) of $\phi$ and equations
(\ref{2.14a}) (\ref{2.14b}) and (\ref{2.18}) to obtain
\begin{flalign*}
&\phi(P)\phi^{-}(P)-(z+\beta)\phi^{-}(P)-(\h z-1)\alpha\\
&=\frac{y/2-V_{11}}{V_{12}}\frac{y/2-V_{11}^{-}}{V_{12}^{-}}-
(z+\beta)\frac{ y/2-V_{11}^{-}}{V_{12}^{-}}-(\h z-1)\alpha\\
&=\frac{1}{V_{12}V_{12}^{-}}\Big[(y/2-V_{11})(y/2-V_{11}^{-})-
(z+\beta) V_{12}(y/2-V_{11}^{-})\\
&~~~~-(\h z-1)V_{12}V_{12}^{-}\Big]\\
&=0.
\end{flalign*}
Equations (\ref{3.13a})-(\ref{3.13c}) are clear from the definitions
of $\phi$ and $y.$
Next we use induction to prove (\ref{3.14a}).

\noindent~(i) $n=n_0;$ one easily finds
$$\psi_{2}(P,n_0,n_{0})=\phi(P,n_0)=\psi_{1}(P,n_0,n_{0})\phi(P,n_0).$$
by definition of $\Psi.$

\noindent(ii)$n>n_0$; we assume (\ref{3.14a}) holds for $n=n_0,\ldots, n-1.$
Then $\psi_1,\psi_2$ satisfy
\begin{flalign}\label{3.16}
  \frac{\psi_{2}(P,n,n_{0})}{\psi_{1}(P,n,n_{0})}
=&\frac{(\h z-1)\alpha\phi^{-}(P,n)^{-1}+z +\beta}{\phi^{-}(P,n)}
\frac{\psi_{2}^{-}(P,n,n_0)}{\psi_{1}^{-}(P,n,n_0)}\nonumber\\
=&\frac{(\h z-1)\alpha\phi^{-}(P,n)^{-1}+z +\beta}{\phi^{-}(P,n)}\phi^-(P,n,n_0)\nonumber\\
=&(\h z-1)\alpha\phi^{-}(P,n)^{-1}+z +\beta,
 \end{flalign}
that is,
\eql{3.17}
{\frac{\psi_{2}(P,n,n_{0})}{\psi_{1}(P,n,n_{0})}\phi^{-}(P,n)-(z+\beta)\phi^{-}(P,n)-(\h z-1)\alpha=0.}
Comparing (\ref{3.17}) with (\ref{3.12}) then
yields (\ref{3.14a}) for $n\geq n_0, ~n\in\mathbb{N}.$

{\renewcommand\baselinestretch{1.3}\selectfont
\noindent (iii)$n<n_0$; analogous proof with (ii).\\
\noindent The definition of $\phi$ (cf. (\ref{3.11b})) implies \par}
\eql{3.18}{\psi_{1}(P,n,n_{0})=\psi_{1}^{-}(P,n,n_{0})\phi^{-}(P,n)}
and hence
\eql{3.19}{
\psi_{1}(P,n,n_{0})=\psi_{2}^{-}(P,n,n_{0}),
}
which follows from (\ref{3.14a}) and (\ref{3.18}).
The definition of $\psi_2$ (cf. (\ref{3.11c})) implies
\begin{flalign}
&\psi_{2}(P,n,n_{0})=\left(\frac{(\h z-1)\alpha}{\phi^{-}(P,n)}+z+\beta\right)\psi_{2}^{-}(P,n,n_{0})\nonumber\\
&=(\h z-1)\alpha\psi_{1}^{-}(P,n,n_{0})+(z+\beta)\psi_{2}^{-}(P,n,n_{0}),\label{3.20}
\end{flalign}
where we use (\ref{3.14a}) again. Then
equation (\ref{3.14b}) follows from
(\ref{3.19}) and (\ref{3.20}).
Property (\ref{3.14c}) is an immediate consequence of (\ref{3.14a})
and the definition of $\phi.$ Finally,
Equations (\ref{3.14d})-(\ref{3.14f}) follow from (\ref{3.13a})-(\ref{3.13c})
,the definition of $\psi_1$ (cf. (\ref{3.11b})) and (\ref{3.14a}).\qed \vspace{0.4cm}

~~Combining the polynomial recursion approach in the
section \ref{section2} with (\ref{3.4})
yields the following trace formula, which means $f_{\ell},g_{\ell}$ can be
expressed by the symmetric functions of the zeros
$\mu_{j}$ of $V_{12}$. For simplicity, we only show one of them.
\newtheorem{lem3.2}[lem3.1]{Lemma}
\begin{lem3.2}\label{lemma3.2}\emph{
 Suppose that $\al,\be$ satisfy the $p$th stationary Ruijsenaars-Toda
 system (\ref{2.12}). Then,
 \eql{3.21}{-\h(\alpha+\alpha^+)-\beta+\delta_1=-\sum_{j=1}^{p}\mu_j.}}
\end{lem3.2}
\noindent{\bf Proof.} Relation (\ref{3.21}) are proved by comparison of powers
of $z$ equating the corresponding expression (\ref{3.4}) for $V_{12}$ with
that in (\ref{2.6a}) and with (\ref{2.3}) taken into account.\qed\vspace{0.2cm}

Next we turn to asymptotic properties of $\phi$ and $\psi_{1}$
in a neighborhood of $P_{\infty\pm}$ and $P_\h.$
The asymptotic behavior of $\Psi_{2}$ is derived naturally from (\ref{3.14a}).
This is a crucial step to construct the stationary algebro-geometric
solutions of stationary Ruijsenaars-Toda hierarchy.
\newtheorem{lem3.3}[lem3.1]{Lemma}
\begin{lem3.3}\label{lemma3.3}\emph{
Suppose that $\alpha,\beta$ satisfy the $p$-th stationary Ruijsenaars-Toda
system (\ref{2.12}).
Moreover, let $P=(z,y)\in\mathcal{K}_{p}\backslash\{P_{\infty\pm},P_{\h}\},~
(n,n_{0})\in\mathbb{Z}\times\mathbb{Z}$. Then, $\phi$ has the asymptotic
behavior
\begin{eqnarray}
\phi(P)&\underset{\zeta\rightarrow 0}{=}&
\begin{cases}
\zeta^{-1}+(\beta-\h\alpha)+O(\zeta),&P\rightarrow P_{\infty-}, \cr
\h\alpha^++(\h^2\alpha^+\alpha+\alpha-\h\alpha\beta)\zeta+O(\zeta^{2}),
&P\rightarrow P_{\infty+}, \cr
\end{cases}~~\zeta=1/z ,\nonumber\\
&&\label{3.22a}\\
 \phi(P)&\underset{\zeta\rightarrow 0}{=}&~~
\frac{\h\alpha^+}{\h+\beta^+}\zeta+O(\zeta^{2}),~~P\rightarrow P_{\h},
~~\zeta=z-1/\h,\label{3.22b}
\end{eqnarray}
Accordingly,
the component $\psi_1$ of the Baker-Akhiezer vector $\Psi$
have the asymptotic behavior
\begin{eqnarray}\label{3.24}
\psi_1(P,n,n_0)&\underset{\zeta\rightarrow 0}{=}&
\begin{cases}
\zeta^{n_0-n}\left(1+O(\zeta)\right),& P\rightarrow P_{\infty-},\cr
\Gamma\left(h\alpha^+\right)\left(1+O(\zeta)\right),&
P\rightarrow P_{\infty+}, \cr
\end{cases}~~\zeta=1/z ,\nonumber\\
&&\label{3.24a}\\
\psi_1(P,n,n_0)&\underset{\zeta\rightarrow 0}{=}&~
\Gamma\left(\frac{h\alpha^+}{h+\beta^+}\right)\zeta^{n-n_0}
\left(1+O(\zeta)\right),\nonumber\\
&~~&~~~~~~~~~~~~~~~~~~~~~~~~~~~~~P\rightarrow P_{\h},~~\zeta=z-1/\h.\label{3.24b}
\end{eqnarray}
The divisor $(\psi_1)$ of $\psi_1$ is given by
\eql{3.25}
{(\psi_{1}(\cdot,n,n_{0}))=\mathcal{D}_{\underline{\hat{\mu}}(n)}
-\mathcal{D}_{\underline{\hat{\mu}}(n_{0})}
+(n-n_{0})(\mathcal{D}_{P_{\h}}-\mathcal{D}_{P_{\infty-}}).}
}
\end{lem3.3}
\noindent{\bf Proof.} The existence of the asymptotic expansions
of $\phi$ in terms of
the local coordinate $\zeta=1/z$ near $P_{\infty\pm}$,
respectively, $\zeta=z-1/\h$ near $P_\h$ is clear from
the explicit form of $\phi$ in (\ref{3.9a}) and (\ref{3.9b}).
Insertion of the polynomials
(\ref{2.6a}) and (\ref{2.6b}) into (\ref{3.9a}) and (\ref{3.9b})
then yields the explicit expansions coefficients in (\ref{3.22a})
and (\ref{3.22b}). Alternatively, and more efficiently, one can
insert each of the following asymptotic expansions
\eqnarrayl{3.26}
{&&\phi\underset{\zeta\rightarrow 0}{=}\phi_{-1}\zeta^{-1}+\phi_{0}+ O(\zeta),\nonumber\\
 &&\phi\underset{\zeta\rightarrow 0}{=}\phi_{0}+\phi_{1}\zeta+ O(\zeta^2),\\
&&\phi\underset{\zeta\rightarrow 0}{=}\phi_{1}\zeta+\phi_2\zeta^2+O(\zeta^3) \nonumber
 }
into the Riccati-type equation (\ref{3.12}) and, upon comparing coefficients of powers
of $\zeta$, which determines the expansion coefficients $\phi_k$ in (\ref{3.26}),
one concludes (\ref{3.22a}) and (\ref{3.22b}). Expansions (\ref{3.24a})
and (\ref{3.24b}) is an immediate consequence of (\ref{3.14a}), (\ref{3.22a})
and (\ref{3.22b}).
Finally, expression (\ref{3.25}) follows from (\ref{3.10}) and (\ref{3.11b}).
\qed\vspace{0.4cm}

\newtheorem{lem3.5}[lem3.1]{Lemma}
\begin{lem3.5}\label{lem3.5}\emph{
  Suppose that $\alpha,\beta$ satisfy (\ref{3.1}) and
  the $p$th stationary RT system (\ref{2.12}).
  Moreover, assume hypothesis (\ref{2.19})
  and (\ref{3.2}) and let $n\in\mathbb{Z}.$
  Let $\mathcal{D}_{\hat{\underline{\mu}}},~\underline{\hat{\mu}}
  =\{\hat{\mu}_1,\ldots,\hat{\mu}_p\}$ be the Dirichlet divisor
  of degree $p$ associated with $\al,\be,$ and $\phi$
  defined according to (\ref{3.9a}) and (\ref{3.9b}),
  that is
  \eqn{\hat{\mu}(n)=\left(\mu_j(n),-2V_{11}(\mu_j(n),n)
  \right)\in\mathcal{K}_p,~~j=1,\ldots,p.}
   Then $\mathcal{D}_{\underline{\hat{\mu}}(n,t_r)}$ is nonspecial for
   all $n\in\mathbb{Z}$.}
\end{lem3.5}
\noindent{\bf Proof.}
The divisor
$\mathcal{D}_{\underline{\hat{\mu}}(n)}$ is nonspecial if and
only if $\{\hat{\mu}_1(n),\dotsi,\hat{\mu}_p(n)\}$ contains one
pair of $\{\hat{\mu}_j,\hat{\mu}^*_j(n)\}$. Hence,
$\mathcal{D}_{\underline{\hat{\mu}}(n)}$
is nonspecial as long as the projection $\mu_j$ of $\hat{\mu}_j$ are mutually distinct,
$\mu_j(n)\neq\mu_k(n)$ for $j\neq k$. If two or more projection coincide for some $n_0\in\mathbb{Z}$,
for instance,$$\mu_{j_1}(n_0)=\dotsi=\mu_{j_k}(n_0)=\mu_0,\quad k>1,$$
then there are two cases in the
following associated with $\mu_0$.\\
(i) $\mu_0\notin \{E_0, E_1,\dotsi,E_{2p+1}\}$;
we have $V_{11}(\mu_0,n_0)\neq 0$ and
$\hat{\mu}_{j_1}(n_0),\ldots,$ $\hat{\mu}_{j_k}(n_0)$ all meet in the same
sheet. Hence no special divisor
can arise in this manner. \\
(ii)$\mu_0\in \{E_0, E_1,\dotsi,E_{2p+1}\}$;
We assume $\mu_0=E_0$ without loss of generality.
One concludes
$V_{12}(z,n_0)\underset{z\rightarrow E_{0}}{=}O\left((z-E_{0})^2\right)$
and $V_{11}(E_{0},n_0)=0$.
Hence
 \[R_{2p+2}(z,n_0)=-V_{11}^2-(\h z-1)\alpha^+V_{12}V_{12}^+=O\left((\lambda-E_{0})^2\right).\]
This conclusion contradict with the hypothesis (\ref{3.2})
that the curve is nonsingular.
As a result, we have $k=1$ and  $\hat{\mu}_j,~j=1,\ldots,p$ are pairwise distinct.
Then we have completed the proof.
\qed\vspace{0.4cm}

Next, we shall provide an explicit representation of $\phi,\psi_1,\al$ and $\be$
in terms Riemann theta function associated with $\cur.$\vspace{0.4cm}

Let us introduce the holomorphic differentials $\eta_\ell(P)$ on
$\cur$ defined by
\eqn{\eta_\ell=\frac{z^{\ell-1}dz}{y},~~\ell=1,\ldots,p}
and choose an appropriate fixed homology basis
$\{a_j, b_j\}_{j=1}^{r-2}$ on $\cur$
in such a way that the intersection matrix of cycles satisfies
$$a_j\circ b_k=\delta_{j,k},\quad a_j\circ a_k=0,\quad b_j\circ b_k=0,\quad j,k=1,\ldots, r-2. $$
Define an invertible matrix $C \in GL(p, \mathbb{C})$ as
follows
\begin{equation}\label{3.27}
          \begin{split}
        & C=(C_{j,k})_{p \times p}, \quad C_{j,k}=
           \int_{a_k} \eta_j, \\
        &  \underline{c}(k)=(c_1(k),\ldots, c_{p}(k)), \quad
           c_j(k)=(C^{-1})_{j,k},
           \end{split}
       \end{equation}
 and the normalized holomorphic differentials
        \begin{equation}\label{3.28}
          \omega_j= \sum_{\ell=1}^{p} c_j(\ell)\eta_\ell, \quad
          \int_{a_k} \omega_j = \delta_{j,k}, \quad
          \int_{b_k} \omega_j= \Gamma_{j,k}, \quad
          j,k=1, \ldots ,p.
        \end{equation}
    One can see that the matrix $\Gamma=(\Gamma_{i,j})_{p\times p}$ is symmetric, and it has a
    positive-definite imaginary part.

    Next, choosing a convenient base point $Q_0 \in
     \mathcal{K}_{p} \setminus \{P_{\infty\pm},P_\h\}$, the vector of Riemann
     constants $\underline{\Xi}_{Q_0}$ is given by (A.45) \cite{m}, and the Abel maps
      $\underline{A}_{Q_0}(\cdot) $ and
      $\underline{\alpha}_{Q_0}(\cdot)$ are defined by
         \begin{eqnarray*}
           \underline{A}_{Q_0}:\mathcal{K}_{p} \rightarrow
           J(\mathcal{K}_{p})&=&\mathbb{C}^{p}/L_{p},
         \end{eqnarray*}
         \begin{eqnarray*}
          P \mapsto \underline{A}_{Q_0} (P)&=& (A_{Q_0,1}(P),\ldots,
           A_{Q_0,p} (P)) \\
           &=&\left(\int_{Q_0}^P\omega_1,\ldots,\int_{Q_0}^P\omega_{p}\right)
           (\mathrm{mod}~L_{p}),
         \end{eqnarray*}
     and
         \begin{eqnarray*}
          && \underline{\alpha}_{Q_0}:
          \mathrm{Div}(\mathcal{K}_{p}) \rightarrow
          J(\mathcal{K}_{p}),\\
          &&~~~~~\qquad \mathcal{D} \mapsto \underline{\alpha}_{Q_0}
          (\mathcal{D})= \sum_{P\in \mathcal{K}_{p}}
           \mathcal{D}(P)\underline{A}_{Q_0} (P),
         \end{eqnarray*}
    where $L_{p}=\{\underline{z}\in \mathbb{C}^{p}|
           ~\underline{z}=\underline{N}+\Gamma\underline{M},
           ~\underline{N},~\underline{M}\in \mathbb{Z}^{p}\}.$

    For brevity, define the function
      $\underline{z}:\mathcal{K}_{p} \times
      \sigma^{p}\mathcal{K}_{p} \rightarrow \mathbb{C}^{p}$ by\footnote{$\sigma^{p}\mathcal{K}_{p}$=
      $\underbrace{\mathcal{K}_{p}\times\ldots\times\mathcal{K}_{p}}_{p}.$}
     \begin{eqnarray}\label{4.4}
           \underline{z}(P,\underline{Q})&=& \underline{\Xi}_{Q_0}
           -\underline{A}_{Q_0}(P)+\underline{\alpha}_{Q_0}
             (\mathcal{D}_{\underline{Q}}), \nonumber \\
           P\in \mathcal{K}_{p},\,~
           \underline{Q}&=&(Q_1,\ldots,Q_{p})\in
           \sigma^{p}\mathcal{K}_{p},
         \end{eqnarray}
     here $\underline{z}(\cdot,\underline{Q}) $ is
     independent of the choice of base point $Q_0$.
     The Riemann theta
     function $\theta(\underline{z})$ associated with $\mathcal{K}_{p}$ and the homology is
      defined by
     $$\theta(\underline{z})=\sum_{\underline{n}\in\mathbb{Z}}\exp\left(2\pi i<\underline{n},\underline{z}>+\pi i<\underline{n},\underline{n}\Gamma>\right),\quad \underline{z}\in\mathbb{C}^{p},$$
     where $<\underline{B},\underline{C}>=\overline{\underline{B}}
     \cdot\underline{C}^t=\sum_{j=1}^{r-2}\overline{B}_jC_j$
     denotes the scalar product in $\mathbb{C}^{p}$.
     Let $\omega_{P_{\h}P_{\infty-}}^{(3)}$ be the normal
    differential of the third kind holomorphic on
    $\mathcal{K}_{p}\backslash\{P_{\infty+},P_{0}\}$ with simple poles at
   $P_{\infty-}$, $P_{\h}$ and residues -1 , 1, respectively.
   In particular,
   \eqnarrayl{3.29}{
   \omega_{P_{\h}P_{\infty-}}^{(3)}&=&\frac{y-2V_{11}(\h^{-1},n)}{z-\h^{-1}}\frac{dz}{2y}
   +\frac{1}{2y}\prod_{j=1}^{p}(z-\lambda_j)dz\label{3.29a}\\
   &\underset{\zeta\rightarrow 0}{=}&
   \begin{cases}
   \begin{array}{cc}
   \Big(-\zeta^{-1}+(-\frac{1}{4}\sum_{m=0}^{2p+1}E_m-\frac{1}{2}\h^{-1}\\
   +\frac{1}{2}\sum_{j=1}^p\lambda_j)+O(\zeta)\Big)
   d\zeta,
   \end{array}&P\rightarrow P_{\infty-},\cr\\
   (\zeta^{-1}+O(1))d\zeta,&P\rightarrow P_{\h},
   \end{cases}\label{3.29b}
   }
   where the constants $\{\lambda_j\}_{j=1}^p\in\mathbb{C}$
   are uniquely determined by employing the normalization
   \[\int_{a_j}\omega_{P_{\h}P_{\infty-}}^{(3)}=0,~~j=1,\ldots,p.\]
   The explicit formula (\ref{3.29a}) and (\ref{3.29b})
   then indicate the following asymptotic expansion
   near $P_{\infty-}$ (using the local coordinate $\zeta=1/z$),
   \eqnarrayl{3.30}
   {\exp\,\left(\int_{Q_0}^P\omega_{P_{\h}P_{\infty-}}^{(3)}\right)&\underset{\zeta\rightarrow0}{=}&
   c_0\Big(\zeta^{-1}+(-\frac{1}{4}\sum_{m=0}^{2p+1}E_m-\frac{1}{2}\h^{-1}\nonumber\\
  &+&\frac{1}{2}\sum_{j=1}^p\lambda_j)
   +O(\zeta)\Big),~~P\rightarrow P_{\infty-}}
   where $c_0$ is an integration constant only depending on $\cur.$
   Moreover,
   assume $\eta\in\mathbb{C}$ and $|\eta|<$min$\{|E_0|^{-1},|E_{1}|^{-1},|E_{2}|^{-1},\dotsi,|E_{2p+1}|^{-1}\}$
   and abbreviate $$\underline{E}=(E_{0},E_{1},\dotsi,E_{2p+1}).$$ Then \eql{ap2000}{\left(\prod_{m=0}^{2p+1}(1-E_{m}\eta)\right)^{-1/2}=\sum_{k=0}^{+\infty}\hat{c}_{k}(\underline{E})\eta^{k},}
where $$\hat{c}_{0}(\underline{E})=1,\quad \hat{c}_{1}(\underline{E})=\frac{1}{2}\sum_{m=0}^{2p+1}E_{m},$$
$$\hat{c}_k(\underline{E})=
\sum_{j_0,\dotsi,j_{2p+1}=0,j_0+\dotsi+ j_{2p+1}=k}^{k}\frac{(2j_0)!\dotsi(2j_{2p+1})! }{2^{2k}(j_0!)^2(j_{2p+1}!)^2}E_0^{j_0}\ldots E_{2p+1}^{j_{2p+1}},\quad k\in\mathbb{N},\quad \text{etc.}$$
Similarly,
\eql{ap3000}{\left(\prod_{m=0}^{2p+1}(1-E_{m}\eta)\right)^{1/2}=\sum_{k=0}^{+\infty}{c}_{k}(\underline{E})\eta^{k},}
where $${c}_{0}(\underline{E})=1,\quad {c}_{1}(\underline{E})=-\frac{1}{2}\sum_{m=0}^{2p+1}E_{m},$$ \eqnarrayn{
c_k(\underline{E})=
\sum_{j_0,\dotsi,j_{2p+1}=0,j_0+\dotsi+ j_{2p+1}=k}^{k}\frac{(2j_0)!\dotsi(2j_{2p+1})! E_0^{j_0}\ldots E_{2p+1}^{j_{2p+1}}}{2^{2k}(j_0!)^2\ldots (j_{2p+1}!)^2(2j_0-1)\dotsi (2j_{2p+1}-1)}\\ k\in\mathbb{N},\quad \text{etc.}}\vspace{0.4cm}

   Given these preparations, the theta function representations of $\phi, \psi_1,\al,$
   and $\be$ then read as follows.\vspace{0.3cm}

 \newtheorem{them3.4}[lem3.1]{Theorem}
 \begin{them3.4}\label{the3.5}\emph{
   Suppose that $\al,\be$ satisfy (\ref{3.1}) and the $p$th stationary
   RT system (\ref{2.12}). Moreover, assume hypothesis (\ref{2.19})
   and (\ref{3.2}), and let $P\in\cur\backslash\{P_{\infty\pm},P_{\h}\}$
   and $(n,n_0)\in\mathbb{Z}^2.$ Then for each $n\in\mathbb{Z},$
   $\mathcal{D}_{\hat{\underline{\mu}}(n)}$ is nonspecial.
   Moreover,
   \eqnarrayl{3.31}
   {\phi(P,n)&=&C(n)\frac{\theta(\underline{z}(P,\underline{\hat{\mu}}^{+}(n)))}
   {\theta(\underline{z}(P,\underline{\hat{\mu}}(n)))}\exp\left(\int_{Q_{0}}^{P}
   \omega_{P_{\h}P_{\infty-}}^{(3)}\right),\label{3.31a}\\
   \psi_{1}(P,n,n_{0})&=&C(n,n_{0})\frac{\theta(\underline{z}(P,\underline{\hat{\mu}}(n)))}
   {\theta(\underline{z}(P,\underline{\hat{\mu}}(n_{0})))}
   \exp\left((n-n_{0})\int_{Q_{0}}^{P}\omega_{P_{\h}P_{\infty-}}^{(3)}\right),
   \nonumber\\\label{3.31b}
   }
   where
   \begin{equation}\label{3.32}
  C(n)=c_0^{-1}\frac{\theta(\underline{z}(P_{\infty-},\underline{\hat{\mu}}(n)))}
  {\theta(\underline{z}(P_{\infty-},\underline{\hat{\mu}}^+(n)))}
 \end{equation}
  and
\begin{eqnarray}\label{3.33}
C(n,n_{0})=
\begin{cases}
\prod_{n^{'}=n_{0}}^{n-1}C(n^{'}),&n>n_{0},\cr
1,&n=n_{0},\cr
\prod_{n^{'}=n}^{n_{0}-1}C(n^{'})^{-1},&n<n_{0}.
\end{cases}
\end{eqnarray}
The Abel map linearizes the auxiliary divisor
$\mathcal{D}_{\underline{\hat{\mu}}(n)}$ in the sense that
\begin{equation}\label{3.34}
\underline{\alpha}_{Q_{0}}(\mathcal{D}_{\underline{\hat{\mu}}(n)})
=\underline{\alpha}_{Q_{0}}(\mathcal{D}_{\underline{\hat{\mu}}(n_{0})})
-\underline{A}_{P_{\infty-}}(P_{\h})(n-n_{0})
\end{equation}
and $\alpha, \beta$ are the form of
\begin{equation}\label{3.35}
\alpha^+=\frac{c_1}{\h c_0}
\frac{\theta(\underline{z}(P_{\infty-},\underline{\hat{\mu}}(n)))}{\theta(\underline{z}(P_{\infty-},\underline{\hat{\mu}}^+(n)))}
\frac{\theta(\underline{z}(P_{\infty+},
\underline{\hat{\mu}}^{+}(n)))}{\theta(\underline{z}(P_{\infty+},\underline{\hat{\mu}}(n)))}
\end{equation}
and
\begin{eqnarray}\label{3.36}
 \beta&=&(c_1/c_0)
\frac{\theta(\underline{z}(P_{\infty-},\underline{\hat{\mu}}^-(n)))}{\theta(\underline{z}(P_{\infty-},\underline{\hat{\mu}}(n)))}
\frac{\theta(\underline{z}(P_{\infty+},
\underline{\hat{\mu}}(n)))}{\theta(\underline{z}(P_{\infty+},\underline{\hat{\mu}}^-(n)))}\nonumber\\
&-&\,\frac{1}{4}\,\sum_{m=0}^{2p+1}E_m
\,-\,\frac{1}{2}\h^{-1}
\,+\,\frac{1}{2}\sum_{j=1}^p\lambda_j\nonumber \\
&+&
\frac{\partial}{\partial\omega_{j}}\ln
\Big(\frac{\theta(\underline{z}(P_{\infty-},\underline{\hat{\mu}}^{+}(n))+\underline{\omega})}
{\theta(\underline{z}(P_{\infty-},
\underline{\hat{\mu}}(n)))+\underline{\omega})}\Big)\Big|_{\underline{\omega}=0.}
 \end{eqnarray}
 Here $c_0, c_1\in\mathbb{C}$ are integration constants.
 }

 \end{them3.4}

\noindent{\bf Proof.}
By Lemma \ref{lem3.5}, $\mathcal{D}_{\underline{\hat{\mu}}(n)}$ is nonspecial
and
hence the theta functions defined in this lemma are not identical to zero.
Obviously, by (\ref{3.10}) and Riemann-Roch Theorem \cite{r}
$$\phi(P,n)\frac{\theta(\underline{z}(P,\underline{\hat{\mu}}(n)))}
{\theta(\underline{z}(P,\underline{\hat{\mu}}^{+}(n)))}
\exp\left(-\int_{Q_{0}}^{P}\omega_{P_{\h}P_{\infty-}}^{(3)}\right)$$
is holomorphic function on compact Riemann surface $\mathcal{K}_{p}$
and therefore
it is a constant $C(n)$ related to $n$. Then $\phi(P,n)$ has the form (\ref{3.31a}).
Next we account for
the following Taylor expansion near $P_{\infty-}$ (with local coordinate $\zeta=1/z$),
\eqnarrayl{3.37}
{\frac{\theta(\underline{z}(P,\underline{\hat{\mu}}^{+}(n)))}{\theta(\underline{z}(P,\underline{\hat{\mu}}(n)))}
&\underset{\zeta\rightarrow 0}{=}&\frac{\theta(\underline{z}(P_{\infty-},\underline{\hat{\mu}}^{+}(n)))}{\theta(\underline{z}
(P_{\infty-},\underline{\hat{\mu}}(n)))}\Big(1
+\sum_{j=1}^{p}c_{j}(p)
\nonumber\\&\times&
\frac{\partial}{\partial\omega_{j}}\ln\Big(
\frac{\theta(\underline{z}(P_{\infty-},\underline{\hat{\mu}}^{+}(n))+\underline{\omega})}
{\theta(\underline{z}(P_{\infty-},\underline{\hat{\mu}}(n))+\underline{\omega})}\Big)\Big|_{\underline{\omega}=0}\zeta\nonumber\\
&+&O(\zeta^{2})\Big),~~~~~~~\text{as}~~P\rightarrow P_{\infty-}}
and hence (\ref{3.30}) and (\ref{3.37}) indicate
\eqnarrayl{3.38}
{\phi(P,n)&\underset{\zeta\rightarrow 0}{=}&c_0 C(n)\frac{\theta(\underline{z}(P_{\infty-},
\underline{\hat{\mu}}^{+}(n)))}{\theta(\underline{z}(P_{\infty-},\underline{\hat{\mu}}(n)))}
\Big(\zeta^{-1}\,+\,(-\,\frac{1}{4}\,\sum_{m=0}^{2p+1}E_m
-\frac{1}{2}\h^{-1} \nonumber\\
&+&\frac{1}{2}\sum_{j=1}^p\lambda_j)-
\frac{\partial}{\partial\omega_{j}}\ln
\Big(\frac{\theta(\underline{z}(P_{\infty-},\underline{\hat{\mu}}^{+}(n))+\underline{\omega})}
{\theta(\underline{z}(P_{\infty-},\underline{\hat{\mu}}(n)))+\underline{\omega})}\Big)\Big|_{\underline{\omega}=0}
+O(\zeta)\Big),\nonumber\\&&~~~~~~~~~~~~~~~~~~~~~~~~~~~~~~~~~~~~~~~~~~~~~~~~~~~
 \text{as}~\,P\rightarrow P_{\infty-}.
 }
 A comparison of the coefficients of the asymptotic relations (\ref{3.22a})
 and (\ref{3.38}) then yields
 the following expressions for $C(n)$ and $\be-\h\al,$
 \eqnarrayl{3.39}
 {C(n)&=&c_0^{-1}\frac{\theta(\underline{z}(P_{\infty-},\underline{\hat{\mu}}(n)))}{\theta(\underline{z}(P_{\infty-},\underline{\hat{\mu}}^+(n)))},
 \label{3.39a}\\
 \be-\h\al&=&
 -\,\frac{1}{4}\,\sum_{m=0}^{2p+1}E_m
-\frac{1}{2}\h^{-1}
+\frac{1}{2}\sum_{j=1}^p\lambda_j\nonumber\\
&+&
\frac{\partial}{\partial\omega_{j}}\ln
\Big(\frac{\theta(\underline{z}(P_{\infty-},\underline{\hat{\mu}}^{+}(n))+\underline{\omega})}
{\theta(\underline{z}(P_{\infty-},
\underline{\hat{\mu}}(n)))+\underline{\omega})}\Big)\Big|_{\underline{\omega}=0.}\label{3.39b}
}
Similarly, one finds
the following Taylor expansions near $P_{\infty+}$ (with local coordinate $\zeta=1/z$),
\eqnarrayl{3.40}
{\frac{\theta(\underline{z}(P,\underline{\hat{\mu}}^{+}(n)))}{\theta(\underline{z}(P,\underline{\hat{\mu}}(n)))}
&\underset{\zeta\rightarrow 0}{=}&\frac{\theta(\underline{z}(P_{\infty+},\underline{\hat{\mu}}^{+}(n)))}{\theta(\underline{z}
(P_{\infty+},\underline{\hat{\mu}}(n)))}\Big(1
-\sum_{j=1}^{p}c_{j}(p)
\nonumber\\&\times&
\frac{\partial}{\partial\omega_{j}}\ln\Big(
\frac{\theta(\underline{z}(P_{\infty+},\underline{\hat{\mu}}^{+}(n))+\underline{\omega})}
{\theta(\underline{z}(P_{\infty+},\underline{\hat{\mu}}(n))+\underline{\omega})}\Big)\Big|_{\underline{\omega}=0}\zeta\nonumber\\
&+&O(\zeta^{2})\Big),~~~~~~~\text{as}~~P\rightarrow P_{\infty+}}
and
\eqnarrayl{3.41}
{\phi(P,n)&\underset{\zeta\rightarrow 0}{=}&c_1 C(n)\frac{\theta(\underline{z}(P_{\infty+},
\underline{\hat{\mu}}^{+}(n)))}{\theta(\underline{z}(P_{\infty+},\underline{\hat{\mu}}(n)))}
\Big(1\,+\,c_2\,\zeta
+O(\zeta)\Big),\nonumber\\&&~~~~~~~~~~~~~~~~~~~~~~~~~~~~~~~~~~~~~~~~~~~~
 \text{as}~\,P\rightarrow P_{\infty+}}
 where $c_0,c_1$ are constants arising from the limiting procedure.
A comparison of (\ref{3.22a}) and (\ref{3.41})
then yields
\eqnarrayl{3.42}
{\h\al^+&=&c_1 C(n)\frac{\theta(\underline{z}(P_{\infty+},
\underline{\hat{\mu}}^{+}(n)))}{\theta(\underline{z}(P_{\infty+},\underline{\hat{\mu}}(n)))}\nonumber\\
&=&c_1/c_0
\frac{\theta(\underline{z}(P_{\infty-},\underline{\hat{\mu}}(n)))}{\theta(\underline{z}(P_{\infty-},\underline{\hat{\mu}}^+(n)))}
\frac{\theta(\underline{z}(P_{\infty+},
\underline{\hat{\mu}}^{+}(n)))}{\theta(\underline{z}(P_{\infty+},\underline{\hat{\mu}}(n)))},
}which proves (\ref{3.35}).
By (\ref{3.39b}) and (\ref{3.42}), one concludes (\ref{3.36}).
Finally, (\ref{3.34}) is the immediate consequence of (\ref{3.10}) and Abel's theorem \cite{r,s,m,n}.
\qed\vspace{0.4cm}

We conclude this section with the trivial case $p=0$ excluded thus far.
\newtheorem{exm3.6}[lem3.1]{Example}
\begin{exm3.6}\emph{
Assume $p=0$, $P=(z,y)\in\mathcal{K}_0\backslash\{P_{\infty\pm},P_\h\}$
and $(n,n_0)\in\mathbb{Z}^2$. Then,
\begin{equation*}
V_0^+(z,n)=
\left(
  \begin{array}{cc}
    - z/2-\h\alpha-\beta+ \delta_1/2 & 1\\[0.2cm]
    (\h z-1)\alpha^+&  z/2-\h\alpha^++
    \delta_1/2\\
  \end{array}
\right),
\end{equation*}
\begin{equation*}
    \mathcal{K}_0:\quad \mathcal{F}_{0}(z,y)=y^2-(z-E_0)(z-E_1)=0,~~~E_0,E_1 \in\mathbb{C}.
\end{equation*}
and
$\alpha,\beta $ satisfy
\begin{equation*}
    \begin{array}{cc}
    \h\alpha^++\h\alpha+\beta=-(E_0+E_1)/2,\\[0.4cm]
      \left(\h\alpha^++(E_0+E_1)/4\right)\left(-\h\alpha^++\delta_1/2\right)+\alpha^+ =-E_0E_1/4,
    \end{array}
\end{equation*}
that is,
\begin{eqnarray*}
    &&\alpha=-\frac{E_0+E_1}{2\h}\pm\frac{1}{2}\sqrt{\left(\frac{E_0+E_1}{2h}\right)^2+\left(\frac{E_0-E_1}{2}\right)^2},~~
    \beta=\frac{E_0+E_1}{2}.\nonumber\\
\end{eqnarray*}
Moveover,
\begin{eqnarray*}
     V_{12}&=&1,\quad V_{21}\,=\,(\h z-1)\alpha^+,\\
      V_{11}&=&-z/2-\h\alpha-\beta+ \delta_1/2,\\
     V_{22}&=&z/2-\h\alpha^++ \delta_1/2,\\
 \phi(P,n_0)&=&y/2-\left(-z/2-\h\alpha-\beta+ \delta_1/2\right),\\
  \Psi_1(P,n,n_0)&=&\Big( y/2- (-z/2-h\alpha-\beta+\delta_1/2 )\Big)^{n-n_0}.
\end{eqnarray*}
}
\end{exm3.6}

\section{Time-dependent Algebro-geometric Solutions}

In this section we extend the algebro-geometric analysis of
Section \ref{section3} to the time-dependent Ruijsenaars-Toda hierarchy.

For most of this section we assume the following hypothesis.
\newtheorem{hyp4.1}{Hypothesis}[section]
\begin{hyp4.1}\emph{
(i) Suppose that $\al,\be$ satisfy
\begin{eqnarray}\label{4.1}
&&\alpha(\cdot,t),\beta(\cdot,t)\in\mathbb{C}^{\mathbb{Z}},\quad \alpha(n,t)\neq 0,\quad (n,t)
\in\mathbb{Z}\times\mathbb{R},\nonumber\\
&&\alpha(n,\cdot),\beta(n,\cdot)\in C^1(\mathbb{R}).
\end{eqnarray}
(ii) Assume that the hyperelliptic curve $\cur$ satisfies (\ref{2.19}) and (\ref{3.2}).}
\end{hyp4.1}

The basic problem in the analysis of algebro-geometric solutions of the
RT hierarchy consists in solving the time-dependent $r$th Ruijsenaars-Toda flow
with initial data a stationary solution of the $p$th equation in the hierarchy.
More precisely, given $p\in\mathbb{N}_0,$ consider a solution $\al^{1}(n), \be^{1}(n)$
of the $p$th stationary Ruijsenaars-Toda system s-$\text{RT}_p(\al^{1},\be^{1})=0$
associated with $\mathcal{K}_p$ and a given set of integration constants
$\{\delta_\ell\}_{\ell=0}^{p}\subseteq\mathbb{C}$. Next, let
$r\in\mathbb{N}_0;$ we intend to construct solutions $\al,\be$ of the $r$th Ruijsenaars-Toda flow
$\text{RT}_r(\al,\be)=0$ with $\al(n,t_{0,r})=\al^{1}(n), \be(n,t_{0,r})=\be^{1}(n)$
for some $t_{0,r}\in\mathbb{R}$. To emphasize that the integration
constants in the definitions of the stationary and the time-dependent RT
equations are independent of each other, we indicate this by adding a tilde on
all the time-dependent quantities. Hence we shall employ the notation
$\widetilde{V}_r, \widetilde{\textrm{RT}}_r,\widetilde{V}_{11},\widetilde{V}_{12}, \widetilde{V}_{21},
\widetilde{V}_{22},\tilde{g}_{\ell},\tilde{f}_{\ell},
\tilde{\delta}_{\ell},$  in order to distinguish them from
$V_p,\textrm{RT}_p,{V}_{11},V_{12},V_{21},V_{22},
g_{\ell}, f_{\ell},\delta_{\ell},$ in the following.
In addition, we will follow a more elaborate
notation inspired by Hirota's $\tau$-function approach
and indicate the individual $r$th RLV flow by a separate time variable $t_r\in\mathbb{R}.$

The algebro-geometric initial value problem discussed above can
be summed up
in the form of zero-curvature equation
\begin{equation}\label{4.2}
     U_{t_{r}}(z,t_{r})+U
     (z,t_{r})\widetilde{V}_{r}(z,t_{r})-\widetilde{V}^{+}_{r}(z,t_{r})U(z,t_{r})=0,
\end{equation}
\begin{equation}\label{4.3}
     U(z,t_{r})V_{p}(z,t_{r})-
     V^{+}_{p}(z,t_{r})U(z,t_{r})=0.
\end{equation}
For further reference, we recall the relevant quantities here (cf. (\ref{2.1}), (\ref{2.5})-(\ref{2.6c})):
\eqnarrayl{4.4}
{&&U=\left(
        \begin{array}{ccc}
          0 & 1 \\[0.2cm]
          (\hbar z-1)\alpha & z+\beta \\
           \end{array}
      \right),\label{4.4a}\\
&&V_p=\left(
       \begin{array}{cc}
         V_{11}^- & V_{12}^- \\[0.2cm]
         V^-_{21} & V^-_{22} \\
       \end{array}
     \right), ~~\widetilde{V}_r=\left(
       \begin{array}{cc}
         \widetilde{V}_{11}^- & \widetilde{V}_{12}^- \\[0.2cm]
         \widetilde{V}^-_{21} & \widetilde{V}^-_{22} \\
       \end{array}
     \right),\label{4.4b}}
and
\eqnarrayl{4.5}
{V_{11}&=&\sum_{j=0}^{p+1}g_{p+1-j} z^j+f_{p+2},
~~V_{12}~=~\sum_{j=0}^{p+1}f_j z^{p+1-j}=\prod_{j=1}^p(z-\mu_j),\nonumber\\
V_{21}&=&(\hbar z-1)\alpha^+V_{12}^+
~=~(\hbar z-1)\alpha^+\left(\sum_{j=0}^{p+1}f_j^+z^{p+1-j}\right),
\nonumber\\
V_{22}&=&-\sum_{j=0}^{p+1}g_{p+1-j}z^j,\nonumber\\
\widetilde{V}_{11}&=&\sum_{j=0}^{r+1}\tilde{g}_{r+1-j} z^j+\tilde{f}_{r+2},
~~\widetilde{V}_{12}~=~\sum_{j=0}^{r+1}\tilde{f}_j z^{r+1-j}, \\
\widetilde{V}_{21}&=&(\hbar z-1)\alpha^+\widetilde{V}_{12}^+
~=~(\hbar z-1)\alpha^+\left(\sum_{j=0}^{r+1}\tilde{f}_j^+z^{r+1-j}\right),
 \nonumber\\
\widetilde{V}_{22}&=&-\sum_{j=0}^{r+1}\tilde{g}_{r+1-j}z^j\nonumber}
for fixed $p\in\mathbb{N}_0\backslash\{0\}, r\in\mathbb{N}_0.$
Here $\{\tilde{g}_\ell\}_{\ell=0}^{r+1}, \{\tilde{f}_\ell\}_{\ell=0}^{r+2}, $ and
$\{g_\ell\}_{\ell=0}^{p+1}, \{f_\ell\}_{\ell=0}^{p+2}$ are
defined by (\ref{2.2a})-(\ref{2.2d}) corresponding to different
constants $\tilde{\delta}_\ell$ and $\delta_\ell$, respectively.
Explicitly,
(\ref{2.4}) (cf. (\ref{2.14a}),(\ref{2.14b})) and (\ref{2.15}) are equivalent to
\eqnarrayl{4.6}
{ && V^-_{11}+  V_{11}+(z+\beta) V_{12}=0,\label{4.6a}\\
 &&(\h z-1)\alpha V^-_{12}-(z+\beta)V^-_{11}
 -(\h z-1)\alpha^+V_{12}^++(z+\beta)V_{11}=0,\label{4.6b}\\
 &&\widetilde{V}_{11}+(z+\beta)\widetilde{V}_{12}-\widetilde{V}_{22}^-=0,\label{4.6c}\\
 &&\alpha_{t_r}+\alpha \widetilde{V}^-_{11}+(z+\beta)\alpha \widetilde{V}_{12}-\alpha \widetilde{V}_{22}=0,\label{4.6d}\\
 &&\beta_{t_r}+(\h z-1)\alpha \widetilde{V}^-_{12}+(z+\beta)\widetilde{V}^-_{22}-
 (\h z-1)\alpha^+\widetilde{V}_{12}^+-(z+\beta)\widetilde{V}_{11}=0,\nonumber\\\label{4.6e}
 }
 respectively.
In particular, (\ref{2.18}) holds in the present $t_r$-dependent setting, that is,
\eqnarrayl{4.7}
{R_{2p+2}(z)&=&-V_{11}^{2}(z,n,t_r)-V_{12}(z,n,t_r)V_{21}(z,n,t_r),\nonumber\\
&=&-V_{11}^{2}(z,n,t_r)-(\hbar z-1)\alpha^+V_{12}(z,n,t_r)V_{12}^+(z,n,t_r).\nonumber\\}
Here we emphasize that $R_{2p+2}$ is $t_r$-independence (cf. Lemma \ref{lemma4.3}).

As in the stationary context, we introduce
\begin{flalign}
&\hat{\mu}_{j}(n,t_r)=(\mu_{j}(n,t_r),-2V_{11}(\mu_{j}(n,t_r),n,t_r)),~~j=1,\ldots,p~~n\in\mathbb{Z}, \label{4.8a}\\
&\hat{\mu}_{j}^{+}(n,t_r)=(\mu_{j}^{+}(n,t_r),2V_{11}(\mu_{j}^+(n,t_r),n,t_r)),~~j=1,\ldots,p~~n\in\mathbb{Z},\label{4.8b}
\end{flalign}
and note that the regularity assumptions (\ref{4.1}) on $\al,\be$
imply continuity of $\mu_j$ with respect to $t_r\in\mathbb{R}.$

In analogy to (\ref{3.9a}), (\ref{3.9b}), one defines the
following meromorphic function $\phi(\cdot,n,t_r)$
on $\cur,$
\begin{flalign}
\phi(P,n)=&\frac{ y/2-V_{11}(z,n,t_r)}{V_{12}(z,n,t_r)}\label{4.9a}\\
=&\frac{(\hbar z-1)\alpha^+(n,t_r) V_{12}^+(z,n,t_r)}{y/2+V_{11}(z,n,t_r)},\label{4.9b}\\
~~&P=(z,y)\in\mathcal{K}_{p},~ (n,t_r)\in\mathbb{Z}\times\mathbb{R},\nonumber
\end{flalign}
with divisor $(\phi(\cdot,n,t_{r}))$ of $\phi(\cdot,n,t_{r})$
given by
\eql{4.10}
{(\phi(\cdot,n,t_{r}))=\mathcal{D}_{P_{0}\underline{\hat{\mu}}^{+}(n,t_{r})}
-\mathcal{D}_{P_{\infty-}\underline{\hat{\mu}}(n,t_{r})}.}
The time-dependent Baker-Akhiezer vector is then
defined in terms of $\phi$ by
\begin{flalign}
&\Psi(P,n,n_{0},t_{r},t_{0,r})=
\left(\begin{array}{cc}\psi_{1}(P,n,n_{0},t_{r},t_{0,r})\\[0.2cm]
\psi_{2}(P,n,n_{0},t_{r},t_{0,r})
\end{array}\right),\\
&\psi_{1}(P,n,n_{0},t_{r},t_{0,r})=
\exp\left(\int_{t_{0,r}}^{t_{r}}\left(\widetilde{V}_{11}(z,n_{0},s)+
\widetilde{V}_{12}(z,n_{0},s))\phi(P,n_{0},s)\right)ds\right)\nonumber\\
&\times
\begin{cases}
\prod_{n^{\prime}=n_{0}}^{n-1}\phi(P,n^{\prime},t_r),&n^{\prime}>n_{0},\cr
1,&n^{\prime}=n_{0},\cr
\prod_{n^{\prime}=n}^{n_{0}-1}\phi(P,n^{\prime},t_r)^{-1},&n^{\prime}<n_{0},
\end{cases}\label{zp}\\
&\psi_{2}(P,n,n_{0})=\exp\left(\int_{t_{0,r}}^{t_r}\left(\widetilde{V}_{11}(z,n_0,s)
+\widetilde{V}_{12}(z,n_0,s)\phi(P,n_0,s)\right)ds\right)\nonumber\\
&\times\phi(P,n_{0},t_r)
\times\begin{cases}
\prod_{n^{\prime}=n_{0}+1}^{n-1}\left(\frac{\alpha(n^{\prime},t_r)(\h z-1)}{\phi^{-}(P,n^{\prime},t_r)}
+z+\beta(n^{\prime},t_r)\right),&n^{\prime}>n_{0},\cr
1,&n^{\prime}=n_{0},\cr
\prod_{n^{\prime}=n+1}^{n_{0}}\left(\frac{\alpha(n^{\prime},t_r)(\h z-1)}{\phi^{-}(P,n^{\prime},t_r)}+
z+\beta(n^{\prime},t_r)\right)^{-1},&n^{\prime}<n_{0},
\end{cases}\\
&~~~~~~~~~~~~P=(z,y)\in\mathcal{K}_p\backslash\{P_{\infty\pm},P_0\},\quad (n,t_r)\in\mathbb{Z}\times\mathbb{R}.
\end{flalign}
One observes that
\begin{eqnarray}\label{4.12}
\psi_{1}(P,n,n_{0},t_{r},t_{0,r})=\psi_{1}(P,n_{0},n_{0},t_{r},t_{0,r})\psi_{1}(P,n,n_{0},t_{r},t_{r})\cr
~~~P\in\mathcal{K}_{p}\backslash\{P_{\infty\pm},P_{0}\},(n,n_{0},t_{r},t_{0,r})\in\mathbb{Z}^{2}\times\mathbb{R}^{2}.
\end{eqnarray}

The following lemma records basic properties of $\phi$ and $\Psi$
in analogy to the stationary case discussed in Lemma \ref{lemma3.1}.

\newtheorem{lem4.1}[hyp4.1]{Lemma}
\begin{lem4.1}\emph{
Assume Hypothesis \ref{4.1} and suppose that (\ref{4.2}), (\ref{4.3})
hold. In addition, let $P=(z,y)\in\cur\backslash\{P_{\infty\pm},P_\h\},
(n,n_0,t_r,t_{0,r})\in\mathbb{Z}^2\times\mathbb{R}^2.$ Then $\phi$
satisfies
  \eqnarrayl{4.14}{
  &&\phi(P)\phi^{-}(P)-(z+\beta)\phi^{-}(P)-(\h z-1)\alpha=0,\label{4.13}\\
  &&\phi_{t_{r}}(P)=(-\widetilde{V}_{11}+\widetilde{V}_{22})\phi-
  \widetilde{V}_{12}\phi^2+(\h z-1)\alpha^+\widetilde{V}_{12}^+\label{4.14a0},\\
   &&\phi(P)\phi(P^{*})=-\frac{(\h z-1)\alpha^+ V_{12}^+(z)}{V_{12}(z)},\label{4.14a}\\
   &&\phi(P)+\phi(P^{*})=-2\frac{V_{11}(z)}{V_{12}(z)},\label{4.14b}\\
    &&\phi(P)-\phi(P^{*})=\frac{y}{V_{12}(z)}.\label{4.14c}}
Moreover, assuming $P=(z,y)\backslash\{P_{\infty\pm}, P_\h\},$
then $\Psi$ satisfies
\begin{flalign}
   & \psi_{2}(P,n,n_{0},t_r,t_{0,r})=\psi_{1}(P,n,n_{0},t_r,t_{0,r})\phi(P,n,t_r),\label{4.15a}\\
    & U(z)\Psi^{-}(P)=\Psi(P),\label{4.15b}\\
    &V_p(z)\Psi^{-}(P)=(y/2)\Psi^-(P),\label{4.15c}\\
    &\Psi_{t_{r}}(p)=\tilde{V}_{r}^{+}(z)\Psi(P),\label{4.15c1}\\
    &\psi_1(P,n,n_0,t_r,t_{0,r})\psi_1(P^*,n,n_0,t_r,t_{0,r})=(\h z-1)^{n-n_0}\Gamma(\al^+,n,n_0,t_r)\nonumber\\
    & ~~~~~~~~~~~~~~~~~~~~~~~~~~~~~~~~~~
    \times \frac{V_{12}(z,n,t_r)}{V_{12}(z,n_0,t_{0,r})},\label{4.15d}\\
    &\psi_1(P,n,n_0,t_r,t_{0,r})
    \psi_2(P^*,n,n_0,t_r,t_{0,r})+\psi_1(P^*,n,n_0,t_r,t_{0,r})\psi_2(P,n,n_0,t_r,t_{0,r})\nonumber\\
    &~~~~~~~~~~~~~~~~~~~~~~~~~=-2(\h z-1)^{n-n_0}\Gamma(\al^+,n,n_0,t_r)
    \frac{V_{11}(z,n,t_r)}{V_{12}(z,n_0,t_{0,r})},\label{4.15e}\\
    &\psi_1(P,n,n_0,t_r,t_{0,r})\psi_2(P^*,n,n_0,t_r,t_{0,r})
    -\psi_1(P^*,n,n_0,t_r,t_{0,r})\psi_2(P,n,n_0,t_r,t_{0,r})\nonumber\\
    &~~~~~~~~~~~~~~~~~~~~~~~~~=-(\h z-1)^{n-n_0}\Gamma(\al^+,n,n_0,t_r)\frac{y}{V_{12}(z,n_0,t_{0,r})},\label{4.15f}
    \end{flalign}
    where we used the abbreviation
    \eql{3.15}{\Gamma(f,n,n_0,t_r)=
    \begin{cases}
    \prod_{n^\prime=n_0}^{n-1}f(n^\prime,t_r),&n> n_0,\cr
    1,&n=n_0,\cr
    \prod_{n^\prime=n}^{n_0-1}f(n^\prime,t_r)^{-1},&n< n_0.
    \end{cases}}
    Moreover, as long as the zeros of $\mu_j(n_0,s)$ of $V_{12}(\cdot,n_0,s)$ are
    all simple for all $s\in\Omega$, $\Omega\subseteq\mathbb{R}$,
    is an open interval, $\Psi$ is meromorphic on
    $\mathcal{K}_p\backslash\{P_{\infty\pm},P_\h\}$ for $(n,n_0,t_r)\in\mathbb{Z}\times\Omega^2$.
}
\end{lem4.1}
\noindent{\bf Proof.}
Equations (\ref{4.13}), (\ref{4.14a})-(\ref{4.15c}), (\ref{4.15d})-(\ref{4.15f})
are proved as in the stationary case, see Lemma \ref{lemma3.1}.
Thus, we turn to the proof of (\ref{4.14a0}) and (\ref{4.15c1}):
Differentiating the Riccati-type equation (\ref{4.13})
yields
\eqn{\phi_{t_r}\phi^-+\phi\phi_{t_r}^{-}=\beta_{t_r}\phi^-+(z+\beta)\phi_{t_r}^-+(\h z-1)\alpha_{t_r},
}
that is,
\begin{equation*}
\begin{split}
&-\left(\phi^-+\left(\phi-(z+\beta)\right)S^-\right)\phi_{t_r}\\
&=\left((\h z-1)\alpha\widetilde{V}_{12}^-+(z+\beta)\widetilde{V}_{22}^--
(\h z-1)\alpha^+\widetilde{V}_{12}^+-(z+\beta)\widetilde{V}_{22}\right)\phi^-\\
&+(\h z-1)\left(\alpha\widetilde{V}_{11}^-+(z+\beta)\alpha\widetilde{V}_{12}
-\alpha\widetilde{V}_{22}\right)
\end{split}
\end{equation*}
by using (\ref{4.6d}) and (\ref{4.6e}).
This allows one to calculate the righthand side of (\ref{4.14a0})
\begin{equation}\label{4.18}
\begin{split}
&\left(\phi^-+\left(\phi-(z+\beta)\right)S^-)\right)
\left(\widetilde{V}_{11}\phi+\widetilde{V}_{12}\phi^2-\widetilde{V}_{21}-\widetilde{V}_{22}\phi\right)\\
&=\widetilde{V}_{11}\phi\phi^-+\widetilde{V}_{12}\phi^2\phi^--\widetilde{V}_{21}\phi^-
-\widetilde{V}_{22}\phi\phi^-+(\h z-1)\alpha\widetilde{V}_{11}^-\\
&+(\h z-1)\alpha\widetilde{V}_{12}^-\phi^--\widetilde{V}_{21}^-\phi+(z+\beta)\widetilde{V}_{21}^--\widetilde{V}_{22}^-(\h z-1)\alpha\\
&=\widetilde{V}_{11}\left((\h z-1)\alpha+(z+\beta)\phi^-\right)+\widetilde{V}_{12}\phi\left((\h z-1)\alpha+(z+\beta)\phi^-\right)\\
&-\widetilde{V}_{21}\phi^--\widetilde{V}_{22}\left((\h z-1)\alpha+
(z+\beta)\phi^-\right)+(\h z-1)\alpha\widetilde{V}_{11}^-+(\h z-1)\alpha\widetilde{V}_{12}^-\phi^-\\
&-\widetilde{V}_{21}^-\phi+(z+\beta)\widetilde{V}_{21}^--\widetilde{V}_{22}^-(\h z-1)\alpha.\\
\end{split}
\end{equation}
Hence,
\begin{equation}\label{4.19}
\begin{split}
&\left(\phi^-+\left(\phi-(z+\beta)S^-\right)\right)
\left(-\phi_{t_r}+\widetilde{V}_{11}\phi+\widetilde{V}_{12}\phi^2-\widetilde{V}_{21}-\widetilde{V}_{22}\phi\right)\\
&=-(z+\beta)\widetilde{V}_{22}^-\phi^-+\widetilde{V}_{11}(z+\beta)\phi^-+(z+\beta)^2\widetilde{V}_{12}\phi^-\\
&=0,\\
\end{split}
\end{equation}
using (\ref{4.6c}).
Solving the first-order difference equation (\ref{4.19}) then yields
\eql{4.20}
{\begin{split}&-\phi_{t_r}+\widetilde{V}_{11}\phi+\widetilde{V}_{12}\phi^2
-\widetilde{V}_{21}-\widetilde{V}_{22}\phi\\
&=E(P,t_r)
\times
\begin{cases}
\prod_{n^{\prime}=1}^{n}B(P,n^{\prime},t_r),&n\in\mathbb{N},\cr
1,&n=0,\cr
\prod_{n^{\prime}=0}^{n+1}B(P,n^{\prime},t_r)^{-1},&-n\in\mathbb{N},\cr
\end{cases}
\end{split}
}
where
\eqn{
B(P,n^{\prime},t_r)
=\frac{\phi(P,n^{\prime},t_r)-(z+\beta(n^{\prime},t_r))}{\phi^{-}(P,n^{\prime},t_r)},
~~ (n^{\prime},t_r)\in\mathbb{Z}\times\mathbb{R}}
and $E(\cdot,t_r)$ is some $n$-independent meromorphic function on $\mathcal{K}_p$.
The asymptotic behavior of $\phi(P,n,t_r)$ in (\ref{3.22a}) then yields (for $t_r\in\mathbb{R}$
fixed)
 \eqn{B(P)\underset{
{ \textrm{ $\zeta\rightarrow 0$}} }{=}\frac{-1}
{\h\alpha}\zeta^{-1}+O(1)~~\textrm{as $P\rightarrow P_{\infty+}$}.}
Comparing the order of both sides in (\ref{4.20})
and taking $n>0$ sufficiently large, one finds contradiction
unless $E=0$. This proves (\ref{4.14a0}).
To prove (\ref{4.15c1}) we rewrite
(\ref{4.14a0}) as
\begin{equation}
\begin{split}
&\phi_{t_r}=\left(-\widetilde{V}_{11}+\widetilde{V}^+_{11}+\left(z+\beta^+\right)
\widetilde{V}_{12}^+-\widetilde{V}_{12}\phi+\frac{\left(\h z-1\right)\alpha^+
\widetilde{V}_{12}^+}{\phi}\right)\phi\\
&=\left(-\widetilde{V}_{11}+\widetilde{V}^+_{11}+\left(z+\beta^+\right)\widetilde{V}_{12}^+
-\widetilde{V}_{12}\phi+\widetilde{V}_{12}^+\phi^+-\left(z+\beta^+\right)\widetilde{V}_{12}^+\right)\phi\\
&=\left(\widetilde{V}_{11}^++\widetilde{V}_{12}^+\phi^+-\widetilde{V}_{11}-\widetilde{B}_{12}\phi\right)\phi\\
\end{split}
\end{equation}
Abbreviating
\eqn{
\Delta(n_0,t_r)=\int_{t_{0,r}}^{t_{r}}
\left(\widetilde{V}_{11}(z,n_{0},s)+\widetilde{V}_{12}(z,n_{0},s)\phi(P,n_{0},s)\right)ds}
one computes for $n\geq n_0+1,$
\begin{flalign}
&\psi_{1,t_{r}}=\left(\exp(\Delta)\prod_{n^{\prime}=n_{0}}^{n-1}
\phi(n^{\prime})\right)_{t_{r}}\nonumber\\
&=\Delta_{t_{r}}\psi_{1}+\exp(\Delta)\sum_{n^{\prime}=n_{0}}^{n-1}
\phi_{t_{r}}(n^{\prime})\prod_{n^{\prime\prime}\neq n^{\prime}}\phi(n^{\prime\prime})\nonumber\\
&=\left(\widetilde{V}_{11}(z,n_{0},t_{r})+\widetilde{V}_{12}
(z,n_{0},t_{r})\phi(P,n_{0},t_{r})\right)\psi_{1}\nonumber\\
&+\exp(\Delta)\sum_{n^{\prime}=n_{0}}^{n-1}\Big(\widetilde{V}_{12}^{+}(n^{\prime})\phi^+(n^{\prime})
+\widetilde{V}_{11}^{+}(n^{\prime})\nonumber\\
&-\widetilde{V}_{12}(n^{\prime})\phi(n^{\prime})
-\widetilde{V}_{11}(n^{\prime})\Big)\phi(n^{\prime})\prod_{n^{\prime\prime}\neq
n^{\prime}}\phi(n^{\prime\prime})\nonumber\\
&=(\widetilde{V}_{11}+\widetilde{V}_{12}\phi)\psi_{1}. \label{4.21}
\end{flalign}
The case $n\leq n_0$ is handled analogously.
By (\ref{4.15a}) and (\ref{4.21}),
\begin{flalign}\label{4.22}
\psi_{2,t_r}=&\phi_{t_r}\psi_1+\phi\psi_{1,t_r}\nonumber\\
=&[(-\widetilde{V}_{11}+\widetilde{V}_{12})\phi-\widetilde{V}_{12}\phi^2
+(\h z-1)\al^+\widetilde{V}^+_{12}]\psi_1+\phi[\widetilde{V}_{11}\psi_1+\widetilde{V}_{12}\psi_2]\nonumber\\
=&\widetilde{V}_{21}\psi_1+\widetilde{V}_{22}\psi_2
\end{flalign}
Combining (\ref{4.21}) with (\ref{4.22}) then yields
(\ref{4.15c1}).

That $\psi_1(P,n,n_0,t_0,t_r)$ is meromorphic on
$\mathcal{K}_p\backslash\{P_{\infty\pm}\}$ if
$V_{12}(\cdot,n_0,t_r)$
has only simple zeros distinct from $P_\h$ is a consequence of
(\ref{4.9a})-(\ref{4.10}), (\ref{zp}), (\ref{4.24a})
and of
\eqnarrayn
{&&\widetilde{V}_{12}~\phi\underset{P\rightarrow\hat{\mu}_j(n_0,s)}{=}
\partial_{s}\ln\left(V_{12}(z,n_0,s)\right)+O(1),~~
 \textrm{as $z\rightarrow \mu_j(n_0,s)$}.~~\qed }\vspace{0.4cm}

 Next we consider the $t_r$-dependence of $V_{11},V_{12}，V_{21}.$

\newtheorem{lem4.3}[hyp4.1]{Lemma}
\begin{lem4.3}\label{lemma4.3}\emph{
Assume Hypothesis \ref{4.1} and suppose that (\ref{4.2}), (\ref{4.3})
hold. In addition, let $(z,n,t_r)\in\mathbb{C}\times\mathbb{Z}\times\mathbb{R}.$
Then
\begin{flalign}
V_{12,t_r}=&(\widetilde{V}_{11}-\widetilde{V}_{22})V_{12}-2V_{11}\widetilde{V}_{12},\label{4.24a}\\
V_{11,t_r}=&\widetilde{V}_{12}V_{21}-\widetilde{V}_{21}V_{12},\label{4.24b}\\
V_{21,t_r}=&2V_{11}\widetilde{V}_{21}+(\widetilde{V}_{22}-\widetilde{V}_{11})V_{21},\label{4.24c}
\end{flalign}
In particular, (\ref{4.24a})-(\ref{4.24c}) are equivalent to
\eql{4.24d}{V_{p,t_r}=[\widetilde{V}_r,V_p]}
and the spectral curve $\cur$ defined by (\ref{2.19}) and (\ref{4.7})
is $t_r$-independent.
}
\end{lem4.3}
\noindent {\bf Proof.}
To prove (\ref{4.24a}) one first differentiates equation
(\ref{4.14c})
\eqn{\phi_{t_r}(P)-\phi_{t_r}(P^{*})=-\frac{y V_{12,t_r}}{V^2_{12}(z)}.}
The time derivative of $\phi$ given in (\ref{4.14a0})
and (\ref{4.14b}) yields
\begin{flalign}
\phi_{t_r}(P)-\phi_{t_r}(P^{*})=&(-\widetilde{V}_{11}+\widetilde{V}_{22})(\phi(P)-\phi(P^*))\nonumber\\
&-\widetilde{V}_{12}(\phi(P)+\phi(P^*))(\phi(P)-\phi(P^*))\nonumber\\
=&(-\widetilde{V}_{11}+\widetilde{V}_{22})y/V_{12}+2\widetilde{V}_{12}V_{11}y/V_{12}^2,\nonumber
\end{flalign}
and hence
\eqn{V_{12,t_r}=(\widetilde{V}_{11}-\widetilde{V}_{22})V_{12}-2V_{11}\widetilde{V}_{12}.}
Similarly, starting from (\ref{4.14b})
\eqn{\phi_{t_r}(P)+\phi_{t_r}(P^*)
=-2\frac{V_{11,t_r}}{V_{12}}+2\frac{V_{11}V_{12,t_r}}{V_{12}^2}}
yields
(\ref{4.24b}).
Moreover,
\begin{flalign*}
V_{21,t_r}=&(\h z-1)\al^+_{t_r}V_{12}^++(\h z-1)\al^+V_{12,t_r}^+\\
=&-(\h z-1)V_{12}^+\Big(\al^+\widetilde{V}_{11}+(z+\be^+)\al^+\widetilde{V}_{12}^+\\
&-\al^+\widetilde{V}_{22}^+\Big)+(\h z-1)\al^+(\widetilde{V}_{11}^+V_{12}^+-\widetilde{V}_{22}^+V_{12}^+\\
&-2V_{11}^+\widetilde{V}_{12}^+)\\
=&-V_{21}\widetilde{V}_{11}-(z+\be^+)V_{12}^+\widetilde{V}_{21}+\widetilde{V}_{11}^+-2V_{11}^+\widetilde{V}_{21}\\
=&2V_{11}\widetilde{V}_{21}+(\widetilde{V}_{22}-\widetilde{V}_{11})V_{21},
\end{flalign*}
using (\ref{4.6a}), (\ref{4.6c}) and (\ref{4.6d}).
Finally, by $(\ref{4.24a})-(\ref{4.24c}),$
differentiating (\ref{4.7}) with respect to $t_r$
then yields
\begin{flalign}
R_{2p+2,t_r}=&-2V_{11}V_{11,t_r}-V_{12,t_r}V_{21}-V_{12}V_{21,t_r}=0.
\end{flalign}
\qed\vspace{0.4cm}

Next we turn to the Dubrovin equation for the time variation of
the zeros $\mu_j$ of $V_{12}$ governed by the $\widetilde{\textrm{RT}}_r$
flow.

\newtheorem{lem4.4}[hyp4.1]{Lemma}
\begin{lem4.4}\emph{
Assume Hypothesis \ref{4.1} and suppose that (\ref{4.2}), (\ref{4.3})
hold
on
$\mathbb{Z}\times\mathcal{I}_\mu$ with $\mathcal{I}_\mu\subseteq\mathbb{R}$
an open interval. In addition, assume that the zeros
$\mu_j,\quad j=1,\ldots,p,$ of $V_{12}$ remain distinct on $\mathbb{Z}\times\mathcal{I}_\mu$.
Then $\{\hat{\mu}_j\}_{j=1,\ldots,p}$ defined in (\ref{4.8a}) and (\ref{4.8b})
satisfy the
following first-order system of differential equation on $\mathbb{Z}\times\mathcal{I}_\mu$,
\eql{4.25}
{\mu_{j,t_{r}}=-\widetilde{V}_{12}(\mu_{j})y(\hat{\mu}_j)
\prod_{\begin{smallmatrix}j=1\\k\neq j\end{smallmatrix}}^{p}\left(\mu_{j}-\mu_{k}\right)^{-1},
~~j=1,\ldots,p,
}
with
\eqn{\hat{\mu}_j(n,\cdot)\in C^{\infty}(\mathcal{I}_\mu,\cur),~~j=1,\ldots,p,~n\in\mathbb{Z}.}
}
\end{lem4.4}
\noindent {\bf Proof.}
It suffices to consider (\ref{4.25}) for $\mu_{j,t_r}.$
Using the product representations for $V_{12}$ in
(\ref{4.5}) and employing (\ref{4.8a}) and (\ref{4.24a}),
one computes
\begin{flalign*}
V_{12,t_r}(\mu_j)=&-\mu_{j,t_r}
\prod_{\begin{smallmatrix}j=1\\k\neq j\end{smallmatrix}}^p\left(\mu_j-\mu_k\right)\\
=&-2V_{11}(\mu_j)\widetilde{V}_{12}(\mu_j)=y(\hat{\mu}_j)\widetilde{V}_{12}(\mu_j),~~j=1,\ldots,p,
\end{flalign*}
proving (\ref{4.25}).\qed\vspace{0.3cm}

Since the stationary trace formulas for $f_\ell$
in terms of symmetric functions of the zeros $\mu_j$
of $V_{12}$ in Lemma \ref{lemma3.2} extend line by line
to the corresponding time-dependent setting, we next record their
$t_r$-dependent analogs without proof. For simplicity
we again confine ourselves to the simplest cases only.\vspace{0.3cm}

\newtheorem{lem4.5}[hyp4.1]{Lemma}
\begin{lem4.5}\emph{
Assume Hypothesis \ref{4.1} and suppose that (\ref{4.2}), (\ref{4.3}) hold.
Then
\eql{4.26}{-\h(\alpha+\alpha^+)-\beta+\delta_1=-\sum_{j=1}^{p}\mu_j.} }
\end{lem4.5}
\vspace{0.3cm}

Next, we turn to the asymptotic expansions of $\phi$ and $\psi_1$
in a neighborhood of $P_{\infty\pm}$ and $P_\h.$\space{0.4cm}

\newtheorem{lem4.6}[hyp4.1]{Lemma}

\begin{lem4.6}\emph{
Assume Hypothesis \ref{4.1} and suppose that (\ref{4.2}), (\ref{4.3}) hold.
Moreover,
let $P=(z,y)\in\cur\backslash\{P_{\infty\pm},P_\h\},(n,n_0,t_r,t_{0,r})
\in\mathbb{Z}^2\times\mathbb{R}^2.$ Then $\phi$ has the asymptotic behavior
\begin{eqnarray}
\phi(P)&\underset{\zeta\rightarrow 0}{=}&
\begin{cases}
\zeta^{-1}+(\beta-\h\alpha)+O(\zeta),&P\rightarrow P_{\infty-}, \cr
\h\alpha^++(\h^2\alpha^+\alpha+\alpha-\h\alpha\beta)\zeta+O(\zeta^{2}),
&P\rightarrow P_{\infty+}, \cr
\end{cases}~~\zeta=1/z ,\nonumber\\
&&\label{4.27a}\\
 \phi(P)&\underset{\zeta\rightarrow 0}{=}&~~
\frac{\h\alpha^+}{\h+\beta^+}\zeta+O(\zeta^{2}),~~P\rightarrow P_{\h},
~~\zeta=z-1/\h.\label{4.27b}
\end{eqnarray}
The component $\psi_1$ of the Baker-Akhiezer vector $\Psi$
has asymptotic behavior
\begin{flalign}
\psi_1(P,n,n_{0},t_{r},t_{0,r})\underset{\zeta\rightarrow 0}{=}&
\exp\left(\pm\frac{1}{2}(t_{r}-t_{0,r})
\sum_{s=0}^{r+1}\tilde{\delta}_{r+1-s}\zeta^{-(s+1)}(1+O(\zeta))\right)\nonumber\\
\times&
\begin{cases}
\zeta^{n_0-n}\left(1+O(\zeta)\right)
 , ~~~~~~~~~~~~
 ~~~~~~~~~~\textrm{as}~~ P\rightarrow P_{\infty-}, \\[0.6cm]
\Gamma(\h\alpha^+)\left(1+O(\zeta)\right)\\
\times
\exp\left(\mathlarger{\int}_{t_{0,r}}^{t_{r}}\left(\sum\limits_{j=0}^{r+1}\tilde{\delta}_{r+1-j}\hat{f}_{j+2}(n_{0},s)\right)ds\right),
\\~~~~~~~~~~~~~~~~~~~
~~~~~~~~~~~~~~~~~~~~~~~~~\textrm{as}~~ P\rightarrow P_{\infty+},
\end{cases}\label{4.28}
\end{flalign}
\begin{flalign}
&\psi_1(P,n,n_{0},t_{r},t_{0,r})\underset{\zeta\rightarrow0}{=}
\Gamma(\frac{\h\alpha^+}{h+\beta^+})\zeta^{n-n_0}(1+O(\zeta))
\exp\left(\int_{t_0}^{t_r}\tilde{V}_{11}(\h,n_0,s)ds\right),\nonumber\\
&~~~~~~~~~~~~~~~~~~~~~~~~~~~~~~~~~~~~~~~~~~~~~~~~~~~~~~~~
~~~~~~~~~~~~~~~~~~~~~~~~~~~~~~P\rightarrow P_\h.\label{4.29}
\end{flalign}
}
\end{lem4.6}
\noindent {\bf Proof.} Since by the definition of $\phi$ in (\ref{4.9a})
and (\ref{4.9b}) the time parameter $t_r$ can be viewed as an additional
but fixed parameter, the asymptotic behavior of $\phi$ remains the same
as in Lemma \ref{lemma3.3}. Similarly, also the asymptotic behavior
of $\psi_1(P,n,n_0,t_r,t_r)$ is derived in an identical fashion to that
in Lemma \ref{lemma3.3}. This proves (\ref{4.28}) for $t_{0,r}=t_r,$
that is,
\begin{flalign}
\psi_1(P,n,n_0,t_r,t_r)\underset{\zeta\rightarrow 0}{=}&
\begin{cases}
\zeta^{n_0-n}\left(1+O(\zeta)\right),& P\rightarrow P_{\infty-},\cr
\Gamma\left(h\alpha^+\right)\left(1+O(\zeta)\right),&
P\rightarrow P_{\infty+}, \cr
\end{cases}~~\zeta=1/z ,\nonumber\\
&\label{4.30}
\end{flalign}
It remain to investigate
\eql{4.31}
{\psi_1(P,n,n_0,t_r,t_{0,r})=\exp\left(\int_{t_{0,r}}^{t_{r}}\left(\widetilde{V}_{11}(z,n_{0},s)+
\widetilde{V}_{12}(z,n_{0},s))\phi(P,n_{0},s)\right)ds\right).}
Next, it is convenient to introduce the homogenous representations of $V_{11},V_{12}$
defined by vanishing all of their integration constants $\delta_\ell, $
that is,
\eqnarrayl{4.32}
{&&\widehat{V}_{11}=\sum_{j=0}^{p+1}\hat{g}_{p+1-\ell}z^\ell+\hat{f}_{r+2}
=V_{11}|_{\delta_0=1,\delta_j=0,j=1,\ldots,p+1},~p\in\mathbb{N},\nonumber\\
&&\widehat{V}_{12}=\sum_{j=0}^{p+1}\hat{f}_{p+1-\ell}z^\ell
=V_{12}|_{\delta_0=1,\delta_j=0,j=1,\ldots,p+1},~p\in\mathbb{N}.\label{4.32a}}
In order to avoid confusion about notation, we relabel $V_{11}, V_{12}$
by $V_{11}^{(p)},V_{12}^{(p)}$ to represent the polynomials associated
with the $p$th stationary Ruijsenaars-Toda equation (\ref{2.12}).
Then we have
\eqnarrayl{4.33}
{&&V_{11}=V_{11}^{(p+1)}=\sum_{j=0}^{p+1}\delta_{p+1-j}\widehat{V}_{11}^{(j)},~~
V_{12}=V_{12}^{(p+1)}=\sum_{j=0}^{p+1}\delta_{p+1-j}\widehat{V}_{12}^{(j)},\nonumber\\
 &&\widetilde{V}_{11}=\widetilde{V}_{11}^{(r+1)}=\sum_{j=0}^{r+1}\tilde{\delta}_{r+1-j}
 \widehat{V}_{11}^{(j)},~~
 \widetilde{V}_{12}=\widetilde{V}_{12}^{(r+1)}=\sum_{j=0}^{r+1}\tilde{\delta}_{r+1-j}
 \widehat{V}_{12}^{(j)}.\nonumber\\
 &&\label{4.33a}}
 Focusing on the homogeneous first, one computes as $P\rightarrow P_{\infty\pm},$
 \begin{flalign*}
 \widehat{V}_{11}^{(j)}+\widehat{V}_{12}^{(j)}\phi
=&\widehat{V}_{11}^{(j)}+\widehat{V}_{12}^{(j)}\frac{y/2-V_{11}^{(p)}}{V_{12}^{(p)}}\\
=&\widehat{V}_{11}^{(j)}+\widehat{V}_{12}^{(j)}\frac{1/2-V_{11}^{(p)}/y}{V_{12}^{(p)}/y}\\
=&\sum_{\ell=0}^{j+1}\hat{g}_{j+1-\ell}z^\ell+\hat{f}_{j+2}+\Big(\sum_{\ell=0}^{j+1}
\hat{f}_{j+1-\ell}z^\ell\Big)\frac{1/2-V_{11}^{(p)}/y}{V_{12}^{(p)}/y}\\
=&\pm\frac{1}{2}\zeta^{-(j+1)}+\begin{cases}\hat{f}_{j+2}+O(\zeta),~~&P\rightarrow
P_{\infty+}\\0+O(\zeta),~~&P\rightarrow P_{\infty-}\end{cases},~~
~~\zeta=1/z.
 \end{flalign*}
One infers from (\ref{4.33a})
\eql{4.34}
{\widetilde{V}_{11}+\widetilde{V}_{12}\phi=\pm\frac{1}{2}\sum_{j=0}^{r+1}\tilde{\delta}_{r+1-j}
\zeta^{-j}+
\begin{cases}
\sum_{j=0}^{r+1}\tilde{\delta}_{r+1-j}\hat{f}_{j+2}+O(\zeta),~&P\rightarrow P_{\infty+},\\
O(\zeta),~&P\rightarrow P_{\infty-}
\end{cases}~\zeta=1/z.}
Insertion of (\ref{4.34}) into (\ref{4.31})
then proves (\ref{4.28}). Finally,
(\ref{4.29}) follows from (\ref{zp}) and (\ref{4.27b}).\qed\vspace{0.4cm}

Next, we turn to the principal result of this section, the representation
of $\phi,\psi_1,\al$ and $\be$ in terms of the Riemann theta function
associated with $\cur,$ assuming $p\in\mathbb{N}$ for the remainder of
this section.\vspace{0.4cm}

Let $\omega_{P_{\infty\pm},q}^{(2)}$ be the normalized differentials
of the second kind with a unique pole at $P_{\infty\pm}$, respectively,
and principal parts
\eql{4.35}
{\omega_{P_{\infty\pm},q}^{(2)}\underset{\zeta\rightarrow0}{=}\left(\zeta^{-2-q}+O(1)
\right)d\zeta,~~P\rightarrow P_{\infty\pm},~~ \zeta=1/z,~q\in\mathbb{N}_{0},
}
with vanishing $a$-periods,
\begin{equation*}
\int_{a_{j}}\omega_{P_{\infty\pm},q}^{(2)}=0,~~ j=1,\ldots,p.
\end{equation*}
Moreover, we define
\eql{4.36}
{
\widetilde{\Omega}_{r}^{(2)}=\frac{1}{2}\left(\sum_{j=1}^{r+1}j
\tilde{\delta}_{r+1-j}\left(\omega_{P_{\infty+},j-1}^{(2)}-\omega_{P_{\infty-},j-1}^{(2)}\right)\right)
}
and
corresponding vector of $b$-periods of $\widetilde{\Omega}^{(2)}_{r}/(2\pi i)$ is then denoted by
$$\underline{\widetilde{U}}_{r}^{(2)}=\left(\widetilde{U}_{r,1}^{(2)},\widetilde{U}_{r,2}^{(2)},
\ldots,\widetilde{U}_{r,p}^{(2)}\right),~~\widetilde{U}_{r,j}^{(2)}=\frac{1}{2\pi i}\int_{b{j}}
\widetilde{\Omega}_{r}^{(2)},~~ j=1,2,\ldots,p.$$
Finally, we abbreviate
\eqnarrayn
{\widetilde{\Omega}_r^{\infty-}=\lim_{P\rightarrow P_{\infty-}}
\left(\int_{Q_0}^P \widetilde{\Omega}_{r}^{(2)}-\frac{1}{2}\sum_{s=0}^{r+1}\tilde{\delta}_{r+1-s}
\zeta^{-(s+1)} \right)
}
and
\eqnarrayn
{\omega_0=\lim_{P\rightarrow P_{\infty-}}\left(\int_{Q_0}^P\omega_{P_\h P_{\infty-}}^{(3)}+\ln \zeta\right).}

\newtheorem{the4.7}[hyp4.1]{Theorem}
\begin{the4.7}\emph{
Assume Hypothesis \ref{4.1} and suppose that (\ref{4.2}), (\ref{4.3}) hold.
Moreover,
let $P=(z,y)\in\cur\backslash\{P_{\infty\pm},P_\h\},(n,n_0,t_r,t_{0,r})
\in\mathbb{Z}^2\times\mathbb{R}^2,$ Then for each $(n,t_r)\in\mathbb{Z}\times\mathbb{R},$
$\mathcal{D}_{\underline{\hat{\mu}}(n,t_r)}$ is nonspecial.
Moreover,
\begin{flalign}
  \phi(P,n)=&C(n,t_r)\frac{\theta(\underline{z}(P,\underline{\hat{\mu}}^{+}(n,t_r)))}
   {\theta(\underline{z}(P,\underline{\hat{\mu}}(n,t_r)))}\exp\left(\int_{Q_{0}}^{P}
   \omega_{P_{\h}P_{\infty-}}^{(3)}\right),\label{4.37a}\\
   \psi_{1}(P,n,n_{0},t_r,t_{0,r})=&C(n,n_{0},t_r,t_{0,r})\frac{\theta(\underline{z}(P,\underline{\hat{\mu}}(n,t_r)))}
   {\theta(\underline{z}(P,\underline{\hat{\mu}}(n_{0},t_{0,r})))}
   \exp\Big((n-n_{0})\int_{Q_{0}}^{P}\omega_{P_{\h}P_{\infty-}}^{(3)}
   \nonumber\\
   &-(t_{r}-t_{0,r})\int_{Q_{0}}^{P}\widetilde{\Omega}_{r}^{(2)}\Big),\label{4.37b}
  \end{flalign}
where
\eql{4.37c}
{ C(n,t_r)=c_0^{-1}\frac{\theta(\underline{z}(P_{\infty-},\underline{\hat{\mu}}(n,t_r)))}
  {\theta(\underline{z}(P_{\infty-},\underline{\hat{\mu}}^+(n,t_r)))}}
and
\eqnarrayl{4.37c1}
{C(n,n_0,t_r,t_{0,r})&=&
\exp\Big((t_r-t_{0,r})\widetilde{\Omega}_r^{\infty-}-(n-n_0)\omega_0
\Big)\nonumber\\
&&\times\frac{\theta(\underline{z}(P_{\infty-},\underline{\hat{\mu}}(n_0,t_{0,r})))}
   {\theta(\underline{z}(P_{\infty-},\underline{\hat{\mu}}(n,t_{r})))}.}
The Abel map linearizes the auxiliary divisor
$\mathcal{D}_{\underline{\hat{\mu}}(n,t_r)}$ in the sense that
\begin{equation}\label{4.66}
\underline{\alpha}_{Q_{0}}(\mathcal{D}_{\underline{\hat{\mu}}(n,t_r)})
=\underline{\alpha}_{Q_{0}}(\mathcal{D}_{\underline{\hat{\mu}}(n_{0},t_{0,r})})
-\underline{A}_{P_{\infty-}}(P_{\h})(n-n_{0})-\underline{\tilde{U}}^{(2)}_r(t_r-t_{0,r})
\end{equation}
and $\alpha, \beta$ are the form of
\begin{equation}\label{4.67}
\alpha^+(n,t_r)=\frac{c_1}{\h c_0}
\frac{\theta(\underline{z}(P_{\infty-},\underline{\hat{\mu}}(n,t_r)))}
{\theta(\underline{z}(P_{\infty-},\underline{\hat{\mu}}^+(n,t_r)))}
\frac{\theta(\underline{z}(P_{\infty+},
\underline{\hat{\mu}}^{+}(n,t_r)))}{\theta(\underline{z}(P_{\infty+},\underline{\hat{\mu}}(n,t_r)))}
\end{equation}
and
\begin{eqnarray}\label{4.68}
 \beta(n,t_r)&=&(c_1/c_0)
\frac{\theta(\underline{z}(P_{\infty-},\underline{\hat{\mu}}^-(n,t_r)))}
{\theta(\underline{z}(P_{\infty-},\underline{\hat{\mu}}(n,t_r)))}
\frac{\theta(\underline{z}(P_{\infty+},
\underline{\hat{\mu}}(n,t_r)))}{\theta(\underline{z}(P_{\infty+},\underline{\hat{\mu}}^-(n,t_r)))}\nonumber\\
&-&\,\frac{1}{4}\,\sum_{m=0}^{2p+1}E_m
\,-\,\frac{1}{2}\h^{-1}
\,+\,\frac{1}{2}\sum_{j=1}^p\lambda_j\nonumber \\
&+&
\frac{\partial}{\partial\omega_{j}}\ln
\Big(\frac{\theta(\underline{z}(P_{\infty-},\underline{\hat{\mu}}^{+}(n,t_r))+\underline{\omega})}
{\theta(\underline{z}(P_{\infty-},
\underline{\hat{\mu}}(n,t_r)))+\underline{\omega})}\Big)\Big|_{\underline{\omega}=0.}
 \end{eqnarray}
Here $c_0, c_1\in\mathbb{C}$ are integration constants.}

\end{the4.7}
\noindent {\bf Proof.} As in Theorem \ref{the3.5} one concludes that
$\phi(P,n,t_r)$ has the form (\ref{4.37a})
and that for $t_{0,r}=t_r, \psi_1(P,n,n_0,t_r,t_{0,r})$ is of the form
\begin{eqnarray*}
\psi_{1}(P,n,n_{0},t_{r},t_{r})=C(n,n_{0},t_{r},t_{r})\frac{\theta(\underline{z}(P,\underline{\hat{\mu}}(n,t_{r})))}
{\theta(\underline{z}(P,\underline{\hat{\mu}}(n_{0},t_{r})))}\cr
\times\exp\left((n-n_{0})\int_{Q_{0}}^{P}\omega_{P_{0}P_{\infty+}}^{(3)}\right).
\end{eqnarray*}
To discuss $\psi_1(P,n,n_0,t_r,t_{0,r})$ we recall (\ref{4.12}), that is,
\eql{4.38}{\psi_{1}(P,n,n_{0},t_{r},t_{0,r})=\psi_{1}(P,n_{0},n_{0},t_{r},t_{0,r})\psi_{1}(P,n,n_{0},t_{r},t_{r}),}
and hence remaining to be studied is
\eql{4.39}
{\psi_{1}(P,n_{0},n_{0},t_{r},t_{0,r})=
\exp\left(\int_{t_{0,r}}^{t_{r}}\widetilde{V}_{11}(z,n_{0},s)+\widetilde{V}_{12}(z,n_{0},s)\phi(P,n_{0},s)\right).
}
Introducing $\hat{\psi}_1(P)$ on $\cur\backslash\{P_{\infty\pm}\}$
by
\eqnarrayl{4.40}
{\hat{\psi}_1&=&C(n_0,n_0,t_r,t_{0,r})\frac{\theta(\underline{z}(P,\underline{\hat{\mu}}(n_0,t_{r})))}
{\theta(\underline{z}(P,\underline{\hat{\mu}}(n_{0},t_{0,r})))}\nonumber\\
&&\times\exp\left(-(t_r-t_{0,r})\int_{Q_0}^P\widetilde{\Omega}_r^{(2)}\right),}
we intend to prove that
\begin{flalign}
\psi_1(P,n_0,n_0,t_r,t_{0,r})=\hat{\psi}_1(P,n_0,t_r,t_{0,r})\nonumber\\
P\in\cur\backslash\{P_{\infty\pm}\},~n_0\in\mathbb{Z},~t_r,t_{0,r}\in\mathbb{R},\label{4.41}
\end{flalign}
for an appropriate choice of the normalization constant $C(n_0,n_0,t_r,t_{0,r})$ in (\ref{4.40}).
We start by noting that a comparison of (\ref{4.28}), (\ref{4.29}), (\ref{4.36}),(\ref{4.37b})
shows that $\psi_1$ and $\hat{\psi}_1$ have the same essential singularities at $P_{\infty\pm}.$
Thus, we turn to the local behavior of $\psi_1$ and $\hat{\psi}_1.$
By (\ref{4.40}) $\hat{\psi}_1$ has zeros and poles
at $\hat{\mu}(n_0,t_r)$ and $\hat{\mu}(n_0,t_{0,r}).$ Similarly, by (\ref{4.39}),
$\hat{\psi}_1$ has zeros and poles only at poles
of $\phi(P,n_0,s)$, $s\in[t_{0,r},t_r]$(resp., $s\in[t_r,t_{0,r}]$).
In the following we temporarily restrict $t_{0,r}$ and $t_r$ to a sufficiently
small nonempty interval $I\subseteq \mathbb{R}$ and pick $n_0\in\mathbb{Z}$
such that for all $s\in I,\mu_j(n_0,s)\neq \mu_k(n_0,s)$ for all $j\neq k, j,k=1,\ldots,p.$
One computes
\begin{flalign}
&~~~~~~~~~\psi_1(P,n_0,n_0,t_r,t_{0,r})\nonumber\\
&~~~~~=\exp\left(\int_{t_0,r}^{t_r}ds\left(\widetilde{V}_{11}
(z,n_0,s)+\widetilde{V}_{12}(z,n_0,s)\frac{ y/2-V_{11}(z,n_0,s)}{V_{12}
(z,n_0,s)}\right)\right)\nonumber\\
&\underset{P\rightarrow\hat{\mu}_j(n_0,s)}{=}\exp\left(\int_{t_{0,r}}^{t_{r}}ds
\left(\frac{\widetilde{V}_{12}\left(\mu_{j}(n_0,s)\right)y(\hat{\mu}_j(n_0,s))}{\left(z-\mu_j(n_0,s)\right)\prod\limits_{
\begin{smallmatrix}k=1\\k\neq j\end{smallmatrix}}^{p}\left(\mu_j(n_0,s)-
\mu_k(n_0,s)\right)}+O\left(1\right)\right)\right)\nonumber\\
&\underset{P\rightarrow\hat{\mu}_j(n_0,s)}{=}
\exp\left(\int_{t_0}^{t_r}ds\left(\frac{-\mu_{j,s}(n_0,s)}{z-\mu_j(n_0,s)}
+O\left(1\right)\right)\right)\nonumber\\
&\underset{P\rightarrow\hat{\mu}_j(n_0,s)}{=}\exp\left(\int_{t_0}^{t_r}ds
\left(\frac{\partial}{\partial s}\ln\left(\mu_j(n_0,s)-z\right)+O(1)\right)\right).\label{4.42}
\end{flalign}
Restricting $P$ to a sufficiently small neighborhood $\mathcal{U}_j(n_0)$ of
$\{\hat{\mu}_j(n_0,s)\in\cur|s\in[t_{0,r},t_r]\subseteq I\}$ such that
$\hat{\mu}_k(n_0,s)\notin \mathcal{U}_j(n_0)$ for all $s\in[t_{0,r},t_r]\subseteq I$
and all $k\in\{1,\ldots,p\}\backslash\{j\},$
(\ref{4.40}) and (\ref{4.42}) imply
\begin{flalign}
&\psi_1(P,n_0,n_0,t_r,t_{0,r})\nonumber\\
=&\begin{cases}
(\mu_j(n_0,t_r)-z)O\left(1\right),& \text{as} \quad P\rightarrow \hat{\mu}_j(n_0,t_r)\neq \hat{\mu}_j(n_0,t_{0,r}),\cr
O\left(1\right),&\text{as}\quad P\rightarrow \hat{\mu}_j(n_0,t_r)=\hat{\mu}_j(n_0,t_{0,r}),\cr
\left(\mu_j(n_0,t_{0,r})-z\right)^{-1}O\left(1\right),&\text{as} \quad P\rightarrow
\hat{\mu}_j(n_0,t_{0,r})\neq \hat{\mu}_j(n_0,t_{r}),
\end{cases}\nonumber\\
&~~~~~~~~~~~~~~~~~~~~~~~~~~~~~~~~~~~~~~~~~~~~~~~
~~~~~~~~~~~~~~~~P=(z,y)\in\cur,\label{4.43}
\end{flalign}
with $O(1)\neq 0.$ Thus $\psi_1$ and $\hat{\psi}_1$ have the same
local behavior and identical essential singularities at $P_{\infty\pm}.$
Hence $\psi_1$ and $\hat{\psi}_1$ coincide up to a multiple constant
(which may depend on $n_0,t_r,t_{0,r}$). By continuity with respect to
divisors this extend to all $n_0\in\mathbb{Z}$ since by hypothesis
$\mathcal{D}_{\underline{\hat{\mu}}(n,s)}$ remain nonspecial for all
$(n,s)\in\mathbb{Z}\times\mathbb{R}.$ Moreover, since by (\ref{4.39}),
for fixed $P$ and $n_0,$ $\psi_1(P,n_0,n_0,\cdot,t_{0,r})$ is entire
in $t_r$(and this argument is symmetric in $t_r$ and $t_{0,r}$),
(\ref{4.41}) holds for all $t_r,t_{0,r}\in\mathbb{R}$(for an appropriate
choice of $C(n_0,n_0,t_r,t_{0,r})$). Together with (\ref{4.38}), this
proves (\ref{4.37b}) for all $(n,t_r),(n_0,t_{0,r})\in\mathbb{Z}\times\mathbb{R}.$

To determine the constant $C(n,n_0,t_r,t_{0,r})$
one compares the asymptotic expansions of $\psi_1(P,n,n_0,t_r,t_{0,r})$
for $P\rightarrow P_{\infty-}$ in (\ref{4.28}) and (\ref{4.37b})
\eqnarrayn{C(n,n_0,t_r,t_{0,r})&=&
\exp\Big((t_r-t_{0,r})\widetilde{\Omega}_r^{\infty-}-(n-n_0)\omega_0
\Big)\\
&&\times\frac{\theta(\underline{z}(P_{\infty-},\underline{\hat{\mu}}(n_0,t_{0,r})))}
   {\theta(\underline{z}(P_{\infty-},\underline{\hat{\mu}}(n,t_{r})))}.}
Finally, (\ref{4.66}) follows from
\begin{flalign*}
&\frac{\partial}{\partial_{t_r}}\underline{\alpha}_{Q_{0},\ell}(\mathcal{D}_{\underline{\hat{\mu}}(n,t_r)})\\
=&\frac{\partial}{\partial_{t_r}}\sum_{j=1}^p\int_{Q_0}^{\hat{\mu}_j(n,t_r)}\omega_\ell\\
=&\sum_{j=1}^p \omega_\ell(\hat{\mu}_j)\mu_{j,t_r}\\
=&\sum_{j=1}^p\left(\sum_{k=1}^p c_\ell(k)\frac{\mu_j^{k-1}}{y(\hat{\mu}_j(n,t_r))}\right)
\left(-\widetilde{V}_{12}(\mu_{j}(n,t_r))y(\hat{\mu}_j(n,t_r))
\prod_{\begin{smallmatrix}k=1\\k\neq j\end{smallmatrix}}^{p}\left(\mu_{j}-\mu_{k}\right)^{-1}\right)\\
=&\sum_{j=1}^p\left(\sum_{k=1}^p c_\ell(k)\frac{\mu_j^{k-1}}{\prod_{k=1,k\neq j}^{p}\left(\mu_{j}-\mu_{k}\right)}\right)\left(-\widetilde{V}_{12}(\mu_{j}(n,t_r))\right)\\
=&-\sum_{k=1}^{p}c_\ell(k)\sum_{j=1}^{p}\frac{\mu_j^{k-1}}{\prod_{k=1,k\neq j}^{p}\left(\mu_{j}-\mu_{k}\right)}\left(\sum_{s=0}^{r+1}\tilde{\delta}_{r+1-s}
\left(\sum_{t=\text{max}\{0,s-p\}}^s\hat{c}_t(\underline{E})\Psi_{s-t}^{(j)}(\underline{\mu})\right)\right)\\
=&-\sum_{k=1}^{p}\sum_{s=0}^{r}c_\ell(k)\tilde{\delta}_{r-s}\hat{c}_{k+s-p}(\underline{E})\\
=&-\underline{\widetilde{U}}^{(2)}_r,
\end{flalign*}
where we use the interpolation representation of $\widetilde{V}_{12}$ in appendix A (cf. (\ref{windows1}))
and
\begin{flalign*}
\omega_j=&\pm\sum_{j=1}^{p}c_{j}(k)\frac{\zeta^{p-j}}{\left(\prod_{m=0}^{2p+1}(1-E_m\zeta)\right)^{\frac{1}{2}}}d\zeta\\
=&\pm\left(\sum_{q=0}^{\infty}\sum_{k=1}^{p}c_j(k)\hat{c}_{k-p+q}(\underline{E})\zeta^q\right)d\zeta,
\\
\tilde{U}_{r,j}^{(2)}=&\frac{1}{2\pi i}\int_{b{j}}\widetilde{\Omega}_{r}^{(2)}\\
=&\frac{1}{2\pi i}\Big[\frac{1}{2}\sum_{s=1}^{r+1}s\tilde{\delta}_{r+1-s}\Big(\int_{b_j}
\omega_{P_{\infty+},s-1}^{(2)}-\int_{b_j}\omega_{P_{\infty-},s-1}^{(2)}\Big)\Big]\\
=&\sum_{s=1}^{r+1}\tilde{\delta}_{r+1-s}\sum_{k=1}^{p}c_l(k)\hat{c}_{k-p+s}(\underline{E}).~~~~~~~~~~~~~~~~~~~~
~~~~~~~~~~~~\qed
\end{flalign*}\vspace{0.4cm}

{\noindent \bf \Large Appendix A: The Lagrange Interpolation
Representation of $\widetilde{V}_{12}(\mu_j(n,t_r))$}\vspace{0.2cm}

 Introducing the notation in \cite{m,n},
\begin{flalign*}
&\Psi_k(\underline{\mu})=(-1)^k\sum_{\underline{\ell}\in\mathcal{S}_k}\mu_{\ell_1}\ldots,\mu_{\ell_k},\\
&\quad \mathcal{S}_k=\{\underline{\ell}=(\ell_1,\ldots,\ell_k)\in\mathbb{N}^k|\ell_1<\ldots<\ell_k\leq p\},~~ k=1,\ldots p,\\
&\Phi_k^{(j)}(\underline{\mu})=(-1)^k\sum_{\underline{\ell}\in\mathcal{\tau}_k^{(j)}}\mu_{\ell_1}\ldots,\mu_{\ell_k},\\ &\mathcal{\tau}_k^{(j)}=\{\underline{\ell}=(\ell_1,\ldots,\ell_k)\in\mathbb{N}^k|\ell_1<\ldots<\ell_k\leq p\quad \ell_m\neq j\},\\
&~~~~~~~~~~~~~~~~~~~~~~~~~~~~~~~~~~~~~~~~~~
k=1,\ldots, p-1,\quad j=1,\ldots,p.
\end{flalign*}
and the formula
\begin{equation}\label{6.3}
\sum_{\ell=0}^{k}\Psi_{k-
\ell}(\underline{\mu})\mu_j^\ell=\Phi_{k}^{(j)}(\underline{\mu}),\quad k=0,\ldots,n,\quad j=1,\ldots,n,
\end{equation}
one finds
$$V_{12}(z)=\sum_{s=0}^{p+1}f_{p+1-s}z^s=\prod_{j=1}^p\left(z-\mu_j\right)=\sum_{\ell=0}^{p}\Psi_{p-\ell}(\underline{\mu})z^\ell$$
and
$$f_\ell=\Psi_{\ell-1}\left(\underline{\mu}\right),~~ \ell=0,\dotsi,p+1.~~
\left( \text{define}~~\Psi_{-1}(\underline{\mu})=0\right)$$
In the case $r<p$,
\begin{align}\label{6.4}
\begin{split}
\widehat{V}_{12}=&\sum_{s=0}^{r+1}\hat{f}_{r+1-s}z^s\\
=&\sum_{s=0}^{r+1}\left(\sum_{k=0}^{\text{min}\{r+1-s,p+1\}}\hat{c}_{r+1-s-k}(\underline{E})f_k\right)z^s\\
=&\sum_{s=0}^{r+1}\left(\sum_{k=0}^{r+1-s}\hat{c}_{r+1-s-k}(\underline{E})f_k\right)z^s\\
=&\sum_{s=0}^{r+1}\left(\sum_{k=0}^{r+1-s}\hat{c}_{r+1-s-k}(\underline{E})\Psi_{k-1}(\underline{\mu})\right)z^s\\
=&\sum_{s=0}^{r+1}\hat{c}_{s}(\underline{E})\sum_{t=0}^{r+1-s}\Psi_{r-s-t}(\underline{\mu})z^t\\
=&\sum_{s=0}^{r}\hat{c}_{s}(\underline{E})\sum_{t=0}^{r-s}\Psi_{r-s-t}(\underline{\mu})z^t.\\
\end{split}
\end{align}
Using (\ref{6.3}), we have
\begin{equation}\label{6.5}
\begin{split}
&\widehat{V}_{12}(\mu_j)=\sum_{s=0}^{r}\hat{c}_{s}(\underline{E})\sum_{t=0}^{r-s}\Psi_{r-s-t}(\underline{\mu})\mu_j^t\\
&=\sum_{s=0}^{r}\hat{c}_{s}(\underline{E})\Psi_{r-s}^{(j)}(\underline{\mu}).\\
\end{split}
\end{equation}
In the case $r>p$,
\begin{align}
\begin{split}
&\widehat{V}_{12}(z) \\
=&\sum_{s=0}^{r+1}\hat{f}_{r+1-s}z^s\\
=&\sum_{s=0}^{r+1}\left(\sum_{k=0}^{\text{min}\{r+1-s,p+1\}}\hat{c}_{r+1-s-k}(\underline{E})f_k\right)z^s\\
=&\sum_{s=0}^{r-p}\sum_{k=0}^{p+1}\hat{c}_{r+1-s-k}(\underline{E})\Psi_{k-1}(\underline{\mu})z^s
+\sum_{s=r-p+1}^{r+1}\sum_{k=0}^{r+1-s}\hat{c}_{r+1-s-k}(\underline{E})\Psi_{k-1}(\underline{\mu})z^s\\
=&\sum_{s=0}^{r-p}\sum_{k=0}^{p+1}\hat{c}_{r+1-s-k}(\underline{E})\Psi_{k-1}(\underline{\mu})z^s
+\sum_{s=r-p+1}^{r+1}\sum_{k=0}^{p+1}\hat{c}_{r+1-s-k}(\underline{E})\Psi_{k-1}(\underline{\mu})z^s\\
=&\sum_{k=0}^{p+1}\sum_{s=0}^{r+1}\hat{c}_{r+1-s-k}(\underline{E})\Psi_{k-1}(\underline{\mu})z^s\\
=&\sum_{s=0}^{r+1}\sum_{k=0}^{p+1}\hat{c}_{r+1-s-k}(\underline{E})\Psi_{k-1}(\underline{\mu})z^s\\
=&\sum_{s=0}^{r+1}\sum_{k=0}^{p+1}\hat{c}_{s}(\underline{E})\Psi_{k-1}z^{r+1-s-k}\\
=&\sum_{s=0}^{r-p}\hat{c}_{s}(\underline{E})\left(\sum_{k=0}^{p+1}\Psi_{k-1}
(\underline{\mu})z^{p+1-k}\right)z^{r-p-s}\\
&+\sum_{s=r-p+1}^{r+1}\hat{c}_{s}(\underline{E})
\left(\sum_{k=0}^{p+1}\Psi_{k-1}(\underline{\mu})z^{r+1-s-k}\right)\\
=&\sum_{s=0}^{r-p}\hat{c}_{s}(\underline{E})\left(V_{12}(z)\right)
z^{r-p-s}+\sum_{s=r-p+1}^{r+1}\hat{c}_{s}(\underline{E})\left(\sum_{k=0}^{r+1-s}\Psi_{k-1}(\underline{\mu})z^{r+1-s-k}\right).\\
\end{split}
\end{align}
Then one finds
\begin{flalign}\label{6.7}
\widetilde{V}_{12}(\mu_j)=&
\sum_{s=r-p+1}^{r+1}\hat{c}_{s}(\underline{E})\left(\sum_{k=0}^{r+1-s}\Psi_{k-1}(\underline{\mu})\mu_j^{r+1-s-k}\right)\nonumber\\
=&\sum_{s=r-p+1}^{r+1}\hat{c}_{s}(\underline{E})\Psi_{r-s}^{(j)}(\underline{\mu}).
\end{flalign}
Combining (\ref{6.5}) with (\ref{6.7}) yields
\begin{equation}
\widehat{V}_{12}(z)=\sum_{s=\text{max}\{0,r-p+1\}}^{r+1}\hat{c}_{s}
(\underline{E})\Psi_{r-s}^{(j)}(\underline{\mu})=
\sum_{s=\text{max}\{0,r-p\}}^{r}\hat{c}_{s}(\underline{E})\Psi_{r-s}^{(j)}(\underline{\mu})
\end{equation}
and hence
\begin{equation}\label{windows1}
\begin{split}
\widetilde{V}_{12}(\mu_j)=&\sum_{s=0}^{r+1}\tilde{\delta}_{r+1-s}\widehat{V}_{12}^{(s)}(\mu_j)\\
=&\sum_{s=0}^{r+1}\tilde{\delta}_{r+1-s}\left(\sum_{t=\text{max}\{0,s-p\}}^{s}\hat{c}_{t}(\underline{E})
\Psi_{s-t}^{(j)}(\underline{\mu})\right).
\end{split}
\end{equation}\vspace{0.3cm}

{\noindent \bf \Large Appendix B: Asymptotic Spectral Parameter Expansions}\vspace{0.4cm}

Next, we turn to asymptotic expansions of various quantities in
the case of the  Ruijsenaars-Toda Hierarchy. Consider a fundamental
system of solutions $\Psi_\pm(z,\cdot)=(\psi_{1,\pm}(z,\cdot),\psi_{2,\pm}(z,\cdot))^\top$
of $U(z)\Psi_\pm^-(z)=\Psi_\pm(z)$
for $z\in\mathbb{C}$ (or in some subdomain of $\mathbb{C}$), with $U$ given by
(\ref{2.1}), such that
\eqn{\text{det}(\Psi_-(z),\Psi_+(z))\neq 0.}
Introducing
\eql{ap1}{\phi_\pm=\frac{\psi_{2,\pm}(z,n)}{\psi_{1,\pm}(z,n)},~~z\in\mathbb{C},~~n\in\mathbb{N},}
then $\phi_\pm$ satisfy the Riccati-type equation
\eql{ap2}
{\phi_\pm\phi^{-}_\pm-(z+\beta)\phi^{-}_\pm-(\h z-1)\alpha=0,}
and one introduces in addition,
\begin{flalign}
&\mathfrak{g}=-\frac{\phi_++\phi_-}{2(\phi_+-\phi_-)},\\
&\mathfrak{f}=\frac{1}{\phi_+-\phi_-}.
\end{flalign}
Using the Riccati-type equation (\ref{ap2})
and its consequences,
\begin{flalign*}
&\phi_+\phi_+^--\phi_-\phi_-^--(z+\beta)(\phi_+^--\phi_-^-)=0,\\
&\phi_-^-\phi_+\phi_+^--\phi_-\phi_-^-\phi_+^-=\alpha(\h z-1)(\phi_-^--\phi_+^-),
\end{flalign*}
one derives the identities
\begin{flalign}
&\mathfrak{g}+\mathfrak{g}^{-}+(z+\beta)\mathfrak{f}=0,\label{ap3}\\
&(\h z-1)\alpha\mathfrak{f}^--(z+\beta)\mathfrak{g}^--(\h z-1)\alpha^+\mathfrak{f}^++(z+\beta)\mathfrak{g}=0,\label{ap4}\\
&\mathfrak{g}^{2}-\alpha^+(\h z-1)\mathfrak{f}\mathfrak{f}^{+}z=1/4,\label{ap5}\\
&\mathfrak{f}=\frac{1}{\phi_+-\phi_-}=\frac{\phi_+^-\phi_-^-}{\alpha(\h z-1)(\phi_-^--\phi_+^-)}.\label{ap6}
\end{flalign}
Moreover, (\ref{ap3})-(\ref{ap6}) also permit one to derive
nonlinear difference equations for $\mathfrak{f}$ and $\mathfrak{g}$ separately,
and one obtains
\begin{flalign}
&(z+\beta)^2-(\mathfrak{g}+\mathfrak{g}^+)(\mathfrak{g}+\mathfrak{g}^-)(\h z-1)\alpha^+\label{ap7}\\
&=\frac{1}{4}(z+\beta)^2,\nonumber\\
&\left((z+\beta)^2\mathfrak{f}+(\h z-1)\alpha\mathfrak{f}^--(\h z-1)\alpha^+
\mathfrak{f}^+\right)^2\nonumber\\
&
-4\mathfrak{f}\mathfrak{f}^+\alpha^+(\h z-1)(z+\beta)^2=(z+\beta)^2.\label{ap8}
\end{flalign}

\noindent {\bf Theorem B.1} Assume (\ref{2.12}), $\text{s-RT}_p(\alpha,\beta)=0,$
 and suppose $P=(z,y)\in\cur\backslash\{P_{\infty\pm}\}.$
 Then $\mathfrak{f},\mathfrak{g}$ has the following convergent expansions
 as $|z|\rightarrow \infty,$

 \begin{equation}\label{ap11}
 \begin{split}
 & \mathfrak{g}\underset{\begin{smallmatrix}
  |\zeta|\rightarrow0
   \end{smallmatrix}}{=}\pm\sum_{\ell=0}^{\infty}\hat{g}_{\ell}\zeta^{\ell},~~\zeta=1/z,\\
 & \mathfrak{f}\underset{\begin{smallmatrix}
  |\zeta|\rightarrow0
   \end{smallmatrix}}{=}\pm\sum_{\ell=0}^{\infty}\hat{f}_{\ell}\zeta^{\ell},~~\zeta=1/z,
   \end{split}
\end{equation}
and simultaneously as $P\rightarrow P_{\infty\pm},$
\begin{equation}\label{ap12}
\begin{split}
&V_{11}/y=\pm\sum_{\ell=0}^{\infty}\hat{g}_{\ell}\zeta^{\ell},~~\zeta\rightarrow0,~~\zeta=1/z,\\
&V_{12}/y=\pm\sum_{\ell=0}^{\infty}\hat{f}_{\ell}\zeta^{\ell},~~ \zeta\rightarrow0,~~\zeta=1/z.
\end{split}
\end{equation}
Moreover, one infers for the $E_m$-dependent summation constants
$\delta_\ell,\ell=0,\ldots,p,$ in $V_{ij} (i,j=1,2)$ that
\begin{equation}\label{ap15}
\delta_\ell=c_{\ell}(\underline{E})~~ \ell=0,\ldots,p
\end{equation}
and
\begin{flalign}
 f_\ell=&\sum_{k=0}^{\ell}c_{\ell-k}(\underline{E})\hat{f}_k,~~\ell=0,\ldots,p+1,\label{ap14}\\
\hat{f}_\ell=&\sum_{k=0}^{\text{min}\{\ell,p+1\}}\hat{c}_{\ell-k}(\underline{E})f_k\nonumber\\
=&\sum_{k=0}^{\text{min}\{\ell,p\}}\hat{c}_{\ell-k}(\underline{E})f_k,~~\ell\in\mathbb{N}_0.\label{ap13}
 \end{flalign}
\noindent {\bf Proof.}
Identifying
\begin{equation}
\Psi_+(z,\cdot)~~\text{with}~~
\Psi(P,\cdot,0)~~\text{and}~~ \Psi_-(z,\cdot) ~~\text{with}~~\Psi(P^{*},\cdot,0),
\end{equation}
and similarly, identifying
\begin{equation}
\phi_+(z,\cdot)~~\text{with}~~\phi(P,\cdot)~~
\text{and}~~ \phi_-(z,\cdot)~~\text{with}~~ \phi(P^{*},\cdot),
\end{equation}
a comparison of (\ref{ap1})-(\ref{ap8})
and the result of Lemma
\ref{lemma3.1} and \ref{lemma3.3} shows that we may
also identify
\eqn
{\mathfrak{g}~~\text{with}~~\pm\frac{V_{11}}{y},~~\mathfrak{f}~~\text{with}~~\pm\frac{V_{12}}{y},}
The sign depending on whether $P$ tends to $P_{\infty\pm}$.
Hence we are only to investigate the asymptotic expansions
of $V_{11}/y$ and $ V_{12}/y$.
Dividing $V_{11}$ and $V_{12}$ by $y$, one obtains
\begin{flalign}
&\frac{V_{11}(z)}{y}=\left(\sum_{k=0}^{\infty}\hat{c}_k(\underline{E})z^{-k}\right)
\left(\sum_{\ell=0}^{p+1}g_\ell z^{-\ell}\right)=\sum_{\ell=0}^{\infty}\check{g}_\ell z^{-l},\label{ap9}\\
&\frac{V_{12}(z)}{y}=\left(\sum_{k=0}^{\infty}\hat{c}_k(\underline{E})z^{-k}\right)
\left(\sum_{\ell=0}^{p+1}f_\ell z^{-\ell}\right)=\sum_{\ell=0}^{\infty}\check{f}_\ell z^{-l},\label{ap10}
\end{flalign}
for some coefficients $\check{g}_\ell,\check{f}_\ell $ to be determined next.
Dividing (\ref{2.14a})and (\ref{2.14b}) by $y$ and inserting the expansions
(\ref{ap9}) and (\ref{ap10}) into the resulting equation then yield the recursion
relations (\ref{2.2a})-(\ref{2.2d})(with $f_\ell$ replaced by $\check{f}_\ell$).
Moreover, plugging (\ref{ap9}) and (\ref{ap10}) into (\ref{2.20a}) and (\ref{2.20b})
then yields
$\check{g}_0=-\frac{1}{2}=\hat{g}_0,~
\check{g}_1=\h\alpha^+=\hat{g}_1,~\check{f}_0=0=\hat{f}_0,~\check{f}_1=1=\hat{f}_1.$
Then
we can inductively to show that
$\check{g}_{\ell}=\mathfrak{M}_\ell(\hat{g}_0,\ldots,\hat{g}_{\ell-1})=\hat{g}_\ell, \check{f}_\ell=\mathfrak{N}_\ell(\hat{f}_0,\ldots,\hat{f}_{\ell-1})=\hat{f}_\ell$,
where $\mathfrak{M}_\ell , \mathfrak{N}_\ell$ are the polynomials
in $(\hat{g}_0,\ldots,\hat{g}_{\ell-1})$ and $(\hat{f}_0,\ldots,\hat{f}_{\ell-1})$,
respectively. This implies (\ref{ap11}) and (\ref{ap12}). A comparison of
coefficients in (\ref{ap10}) then proves (\ref{ap13}).
Next, multiplying (\ref{ap2000}) and (\ref{ap3000}), a comparison of
coefficients of $\eta^{-k}$ yields
\begin{equation}
\sum_{\ell=0}^{k}\hat{c}_{k-\ell}(\underline{E})c_\ell(\underline{E})=\delta_{k,0}=\begin{cases}1&k=0\\0&k\neq 0\end{cases}.
\end{equation}
Hence one computes
\begin{flalign*}
\sum_{k=0}^{\ell}c_{\ell-k}(\underline{E})\hat{f}_k=&\sum_{m=0}^{\ell}c_{\ell-k}(\underline{E})
\left(\sum_{s=0}^{\min\{k,p+1\}}\hat{c}_{k-s}(\underline{E})f_s\right)\\
=&\sum_{k=0}^{\ell}c_{\ell-k}(\underline{E})\left(\sum_{s=0}^{k}\hat{c}_{k-s}(\underline{E})f_s\right)\\
=&\sum_{k=0}^{\ell}\sum_{s=0}^{k}c_{\ell-k}(\underline{E})\hat{c}_{k-s}(\underline{E})f_s\\
=&\sum_{k=0}^{\ell}
\sum_{s=0}^{\ell}c_{\ell-k}(\underline{E})\hat{c}_{k-s}(\underline{E})f_s\\
=&\sum_{s=0}^{k}\sum_{k=0}^{\ell}c_{\ell-k}(\underline{E})\hat{c}_{k-s}(\underline{E})f_s
\\
=&\sum_{s=0}^{\ell}
\left(\sum_{k=s}^{\ell}c_{\ell-k}(\underline{E})\hat{c}_{k-s}(\underline{E})\right)f_s\\
=&f_\ell,~~ \ell=0,\ldots,p+1.
\end{flalign*}
Hence one obtains (\ref{ap14}) and thus (\ref{ap15}).\qed\vspace{0.4cm}

\noindent {\bf \Large Acknowledgments}\vspace{0.2cm}

We are very grateful to Professor F.Gesztesy for his helps to
improve our paper. This work was supported
by grants from the National Science Foundation of China (Project
No.10971031), and the Shanghai Shuguang Tracking Project
(Project No.08G
G01).

\renewcommand{\baselinestretch}{1.2}

\end{document}